\documentclass[journal=rsm]{CUP-JNL-DTM}%

\usepackage{graphicx}
\usepackage{multicol,multirow}
\usepackage{amsmath,amssymb,amsfonts}
\usepackage{mathrsfs}
\usepackage{amsthm}
\usepackage{rotating}
\usepackage{appendix}
\usepackage{ifpdf}
\usepackage[T1]{fontenc}
\usepackage{newtxtext}
\usepackage{newtxmath}
\usepackage{textcomp}
\usepackage{xcolor}
\usepackage{lipsum}
\usepackage[colorlinks,allcolors=blue]{hyperref}
\usepackage{booktabs}
\usepackage{makecell}
\usepackage{listings}
\lstdefinelanguage{Stata}{
  morekeywords={use, sysuse, clear, gen, egen, replace, drop, keep, if, in, by,
    sort, summarize, sum, tabulate, tab, regress, reg, logit, probit, xtset,
    tsset, foreach, forvalues, while, do, local, global, macro, quietly, noisily},
  sensitive=true,
  morecomment=[l]{*},
  morecomment=[l]{//},
  morecomment=[s]{/*}{*/},
  morestring=[b]"
}

\theoremstyle{definition}

\numberwithin{equation}{section}

\jname{Research Synthesis Methods}
\articletype{RESEARCH ARTICLE}
\jyear{2025}
\begin{document}

% some custom commands:
\providecommand{\expect}{\mathrm{E}}             % expectation
\providecommand{\var}{\mathrm{Var}}              % variance
\providecommand{\normaldistn}{\mathrm{Normal}}   % normal distribution
\providecommand{\unifdistn}{\mathrm{Uniform}}    % uniform distribution
\providecommand{\halfnormaldistn}{\mathrm{HN}}    % uniform distribution
\providecommand{\rmk}[1]{{[\textit{#1}]}}
\providecommand{\uisd}{\sigma_\mathrm{u}}
% reporting information fraction: 
\providecommand{\reportingif}{\pi_\star} 
% overall information fraction:
\providecommand{\overallif}{\tilde{\pi}} 

\begin{Frontmatter}

\title[Article Title]{Consistent Bayesian random-effects meta-analysis\\ on subgroup specific effects and interactions}

% There is no need to include ORCID IDs in your .pdf; this information is captured by the submission portal when a manuscript is submitted. 
\author[1]{Renato Panaro}
\author[1]{Christian R\"{o}ver}
\author[1,2,3]{Tim Friede}

\authormark{Renato Panaro, Christian R\"{o}ver, Tim Friede}

\address[1]{\orgdiv{Department of Medical Statistics}, \orgname{University Medical Center G\"{o}ttingen}, \orgaddress{\state{G\"{o}ttingen}, \country{Germany}}.\email{rvpanaro@gmail.com}}
\address[2]{\orgname{DZHK (German Center for Cardiovascular Research)}, \orgdiv{partner site Lower Saxony}, \orgaddress{\state{G\"{o}ttingen}, \country{Germany}}}
\address[3]{\orgname{DZKJ (German Center for Child and Adolescent Health)}, \orgaddress{\state{G\"{o}ttingen}, \country{Germany}}}

\keywords{Hierarchical modeling, subgroup analysis, treatment effect modifiers, heterogeneity, subgroup contribution, aggregation bias}

\keywords[MSC Codes]{\codes[Primary]{62F15}; \codes[Secondary]{62C10, 62P10}}
% 62F15: "Bayesian inference"
% 62C10: "Bayesian problems; characterization of Bayes procedures"
% 62P10: "Applications to biology and medical sciences"

\abstract{Clinical trials commonly report effects on the full study population as well as for patient subgroups. Random-effects meta-analyses of subgroup-specific effects and treatment-by-subgroup interactions might be misaligned, especially when subgroups contribute differently to study populations. We show that, after contribution adjustment, subgroup meta-analysis can recover the same interaction inference regardless of whether interaction or subgroup data are used. Our framework for subgroup-data interaction meta-analysis has four advantages. First, it adjusts for variation in relative subgroup contribution, quantified by the within-trial information fraction. Second, it is robust to prevalence imbalance and variation. Third, it provides a self-contained model-based approach. Fourth, it can incorporate prior information into Bayesian interaction meta-analysis.
The method demonstration uses an example application including 7~trials of disease-modifying therapies in relapsing–remitting multiple sclerosis; the contribution-adjusted model manages to reproduce the results from the interaction meta-analysis while yielding matching estimates for the subgroup effects.
By controlling subgroup contribution while retaining subgroup interpretability, this approach allows for interaction decision-making when published subgroup data are available. Although the proposed model is presented in the Bayesian context, it can also be implemented in frequentist or likelihood frameworks.}
\end{Frontmatter}

%\section*{Impact Statement}
%Some Data journals (DAP, DCE) require an `Impact Statement' section. Comment out this section if it is not required.

% Some math journals (FLO) require a table of contents. Comment out this line if no ToC is needed.
%\localtableofcontents

\section*{Highlights}
\paragraph{What is already known:}
\begin{itemize}
  \item Subgroup and interaction effect estimates tend to align when subgroup prevalences are constant across studies
  \item While explicit specification of \emph{estimators} within a common model may enforce desirable properties in a frequentist framework, a Bayesian approach requires adaptation of the data model
\end{itemize}

\paragraph{What is new:}
\begin{itemize}
  \item We introduce the \emph{Contribution-adjusted Meta-analysis by Subgroup (CAMS)} in the Bayesian framework
  \item We show that meta-analytic subgroup comparison is associated with subgroup contribution patterns (information fractions)
  \item We recommend using the \emph{overall information fraction} when reporting subgroup-specific and overall effect estimates
  \item CAMS can also be applied in a frequentist or likelihood framework
\end{itemize}

\paragraph{Potential impact for RSM readers outside the authors' field:}
\begin{itemize}
  \item We offer a Bayesian alternative for matching subgroup and interaction meta-analysis based on aggregate subgroup-data
  \item Similar approaches may resolve collapsibility issues in related (Bayesian or frequentist) estimation problems
  \item We offer a generic contribution-adjusted framework extendable to other types of outcomes and dimensions, such as multiple comparisons and network meta-analysis
\end{itemize}

\section{Introduction}
\label{sec:Introduction}
In randomised controlled trials, subgroup analyses are carried out to either demonstrate homogeneity of the treatment effect across the study population or to explore potential treatment effect modifiers.
In many applications, these baseline characteristics are binary or dichotomised (e.g., older vs younger, biomarker-positive vs.\ biomarker-negative), and trials report subgroup-specific treatment effects (e.g., treatment effect on biomarker-positive patients).  
Subgroup findings are important for clinical decision-making, motivating careful meta-analytic methods for treatment-covariate interactions~[\cite{DiasEtAl2012c, DoneganEtAl2015, BestEtAl2021, Godolphin2022, WangEtAl2024}]. 

Aggregate-data subgroup analyses can conflate within-trial interaction with between-trial case-mix differences, inducing aggregation bias~[\cite{Berlin2002, Cooper2009, RileyEtAl2020}]. In subgroup meta-analysis, studies may have varying subgroup prevalences that differ from the target population. Such misrepresentation can make these studies leverage points in meta-analysis, creating small-study effects~[\cite{Chaimani2012}]. Subgroup comparisons have previously been considered in frequentist meta-analysis using within-trial aggregate-data methods~[\cite{Godolphin2022}] and approaches that separate within- and across-trial information~[\cite{Hua2017, FisherEtAl2017, Riley2023}]. However, there is no direct analogue for Bayesian meta-analysis of aggregate subgroup data, where compatible subgroup and interaction inference must instead be encoded in the joint model. We therefore introduce a model-based contribution adjustment (Section~\ref{sec:IFadjustedBayesianmeaanalysisbysubgroupCAMS}). The core idea behind the proposed \emph{Contribution-adjusted Meta-analysis by Subgroup (CAMS)} model is to formalize and pool evidence for subgroup differences by quantifying each study's contribution to the subgroup contrast and embedding these \emph{within-trial comparisons} in a Bayesian framework.

A Bayesian approach is particularly useful in meta-analyses with only a small number of studies, especially when between-study heterogeneity is present but difficult to estimate~[\cite{FriedeRoeverWandelNeuenschwander2017a, SchmidCarlinWelton2021}]. This is the setting that motivated our application, and more generally it is the setting in which Bayesian random-effects models with weakly informative priors are especially attractive, because they can regularize heterogeneity parameters and provide coherent uncertainty propagation~[\cite{RoeverEtAl2021}]. Bayesian methods are also well suited to contexts such as two-study meta-analysis, rare-disease and orphan-disease applications, and evidence synthesis with few studies, where data are sparse and prior information can be incorporated naturally within a hierarchical model~[\cite{FriedeRoeverWandelNeuenschwander2017b, KnappHartung2003, GunhanRoeverFriede2020, BenderEtAl2018, RoeverFriede2023b}]. Although the proposed contribution-adjusted approach could also be implemented in frequentist or likelihood-based frameworks, the Bayesian formulation is particularly convenient for the few-study setting considered here.

More specifically, we extend the normal–normal hierarchical model to a bivariate random-effects structure that explicitly captures both within- and between-trial variability, linearly adjusted by subgroup contributions (relative subgroup weight/size within a trial).   By incorporating prior information (e.g., weakly informative normal priors for treatment effects and half-normal priors for variance components), the adjusted model is suitable for Bayesian interaction inference, in particular in few-study settings~[\cite{TurnerEtAl2012, FriedeRoeverWandelNeuenschwander2017b}]. Methodologically, we extend the arm-based formulation by Jackson et al.~[\cite{JacksonEtAl2018}] to the subgroup setting and orthogonalize the main effects and interaction terms. Subgroup contributions are reconstructed from subgroup data by using the reported subgroup differences and their variances,
% enabling its application to any published outcome scale.
enabling its application to a wide range of outcomes that would commonly be analyzed using normal models.
The framework unifies \emph{Meta-Analysis by Subgroup (subgroup-MA)} and \emph{Interaction Meta-analysis (interaction-MA)} within a single likelihood.

This work is organized as follows. Section~\ref{sec:DataandOutcomes} introduces motivating examples and terminology; Section~\ref{sec:Methods} outlines the modeling framework; Section~\ref{sec:Results} presents the findings and Section~\ref{sec:Conclusion} concludes.

\section{Motivating example}
\label{sec:DataandOutcomes}

%\subsection{Multiple sclerosis example application}
%\label{sec:Multiplesclerosisexampleapplication}

Multiple sclerosis (MS) is a chronic neurological disorder characterized by relapses, and relapse rates are commonly used as primary endpoints when evaluating treatments in relapsing MS\@. Heesen et al. (2024) [\cite{Heesen2024}] collated data from published randomized controlled trials to investigate whether disease-modifying therapies appear to act uniformly or differentially across several patient subgroups. Analyses have focused on baseline characteristics such as age and disability status, with evidence of greater benefits in younger and less disabled patients [\cite{HauserEtAl2017}]. From each trial, subgroup-specific effect estimates (e.g., rate ratios or odds ratios) were extracted or derived from published reports.

\begin{figure}
\centering
\includegraphics[width =\textwidth]{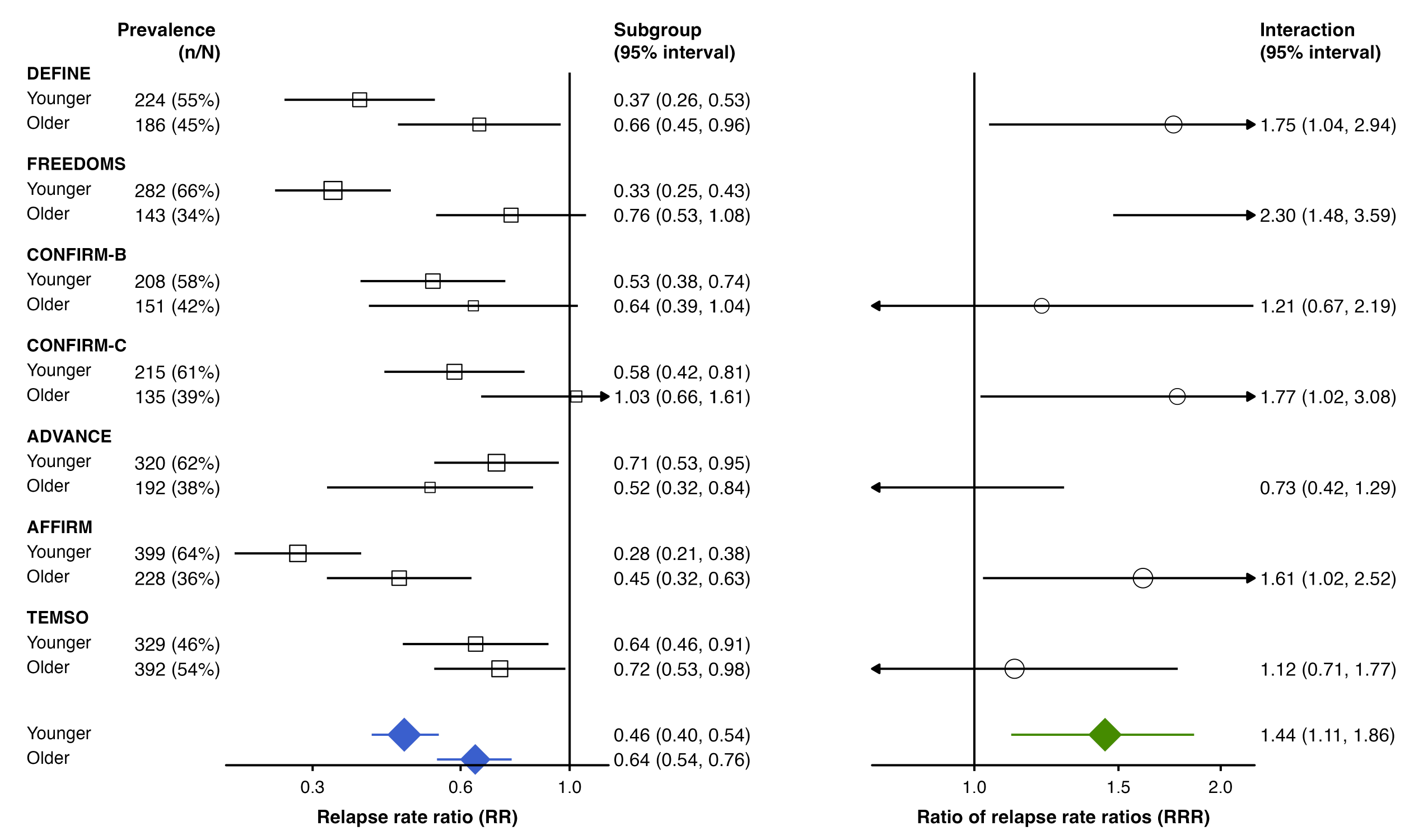}
\caption{Bayesian random-effects meta-analyses of the immunotherapy effect on relapse rates in RRMS stratified by age (age \(< 40\) vs. age \(\ge 40\) years). Following the presentation of subgroup-specific effects and interactions in Fisher et al. ~[\cite{FisherEtAl2017}], the left panel shows subgroup-specific rate ratios (RR), and the right panel shows the treatment-by-subgroup interaction in terms of ratios of RRs (RRR). Posterior medians with \(95\%\) credible intervals for the overall effects are displayed at the bottom}\label{fig:age-forest}
\end{figure}

\begin{figure}
\includegraphics[width = \textwidth]{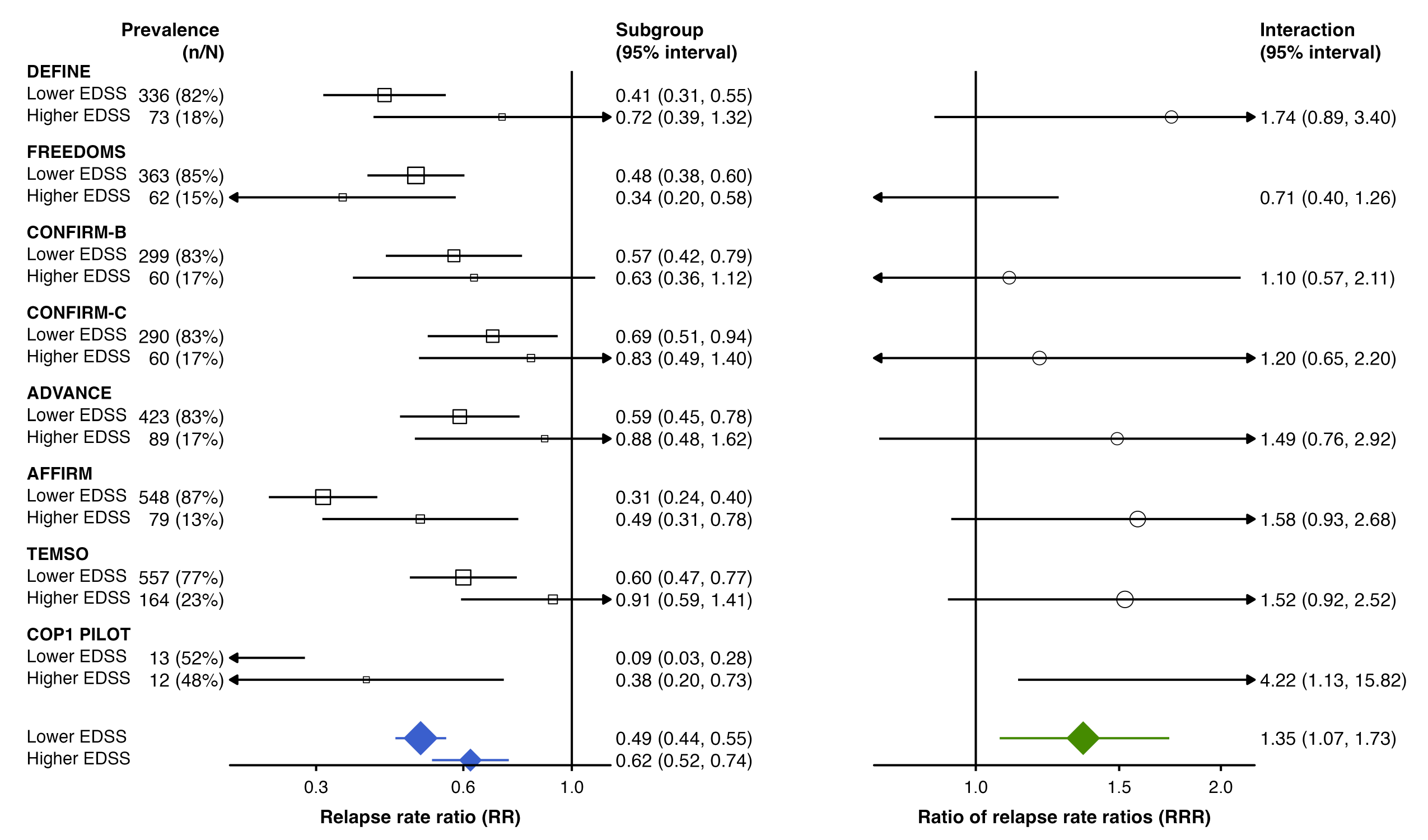}
\caption{Bayesian random effects meta-analyses of immunotherapy effect on relapse rates in RRMS stratified by disease severity subgroups ($\mbox{EDSS} \leq 3.5$ vs. $\mbox{EDSS} > 3.5$). The left panel shows subgroup-specific rate ratios (RR) by baseline EDSS\@. The right panel displays treatment-by-subgroup interactions (ratios of RRs, RRRs) comparing more to less disabled patients. Posterior medians with 95\% credible intervals for overall effects are displayed at the bottom}
\label{fig:edss-forest}
\end{figure}

The forest plots in Figures~\ref{fig:age-forest}--\ref{fig:edss-forest} illustrate subgroup analyses by showing subgroup-specific treatment effects (rate ratios, RR) alongside the estimated treatment-by-subgroup interactions (ratios of RRs, RRR).
Because only a small number of trials were available, we used a Bayesian random-effects model so that weakly informative priors could regularize the between-study heterogeneity~[\cite{WikstenRueckerSchwarzer2016}] and propagate its uncertainty,~[\cite{FriedeRoeverWandelNeuenschwander2017a,FriedeRoeverWandelNeuenschwander2017b}] which is otherwise difficult to estimate reliably in few-study settings~[\cite{RoeverKnappFriede2015, BenderEtAl2018}]. Younger patients and those with lower baseline Expanded Disability Status Scale (EDSS) scores tend to experience greater relative benefit from disease-modifying therapies, as reflected in smaller RRs. These observed effects are likely to be associated, since, due to the progressing nature of MS, younger patients also tend to be less severely diseased. 
Subgroup composition data further illustrate the distribution of participants across subgroups within each trial, with 95\% confidence intervals for the study-specific estimates and overall credible intervals from the pooled Bayesian analyses.
The three summary estimates presented are obtained from three distinct meta-analyses, one conducted within each subgroup and one for the interaction term. These meta-analyses are conducted independently of each other. A coherent joint model for all these components would be desirable.

\section{Standard methods and a new proposal}
\label{sec:Methods}

\subsection{Data structure and notation}
\label{sec:Datastructureandnotation}
We consider the case of two patient subgroups contrasted in a meta-analysis of \(k\) studies. 
In the \(j\)th trial, the treatment effect for subgroup \(A\) is denoted \(y_{Aj}\) and for subgroup \(B\) by \(y_{Bj}\), both measured on continuous scale 
%(e.g., \(\log\mathrm{ORs}\), \(\log\mathrm{RRs}\) or \(\log\mathrm{HRs}\)). 
(such as mean differences, log-ORs, log-RRs or log-HRs).
In matrix notation, the bivariate vector \({y}_j=(y_{Aj},y_{Bj})^\prime\) collects the two endpoints for study~\(j\). 
Each outcome has an associated standard error \(\sigma_{Aj}\) or \(\sigma_{Bj}\), yielding the covariance matrix \({S}_j=\mathrm{diag}(\sigma_{Aj}^2,\sigma_{Bj}^2)\). 
As in standard meta-analytic models, the reported standard errors are treated as fixed and known for synthesis.

For each study $j$, define the within-study contrast
\(
g_j = y_{B j} - y_{A j},
\)
and the study-specific overall effect
\(
m_j = (1-\pi_j)y_{A j} + \pi_j y_{B j},
\)
where $\pi_j$ denotes the \emph{contribution} of group~$B$ in study~$j$.
The contrast~$g_j$ is the primary quantity of interest, while $m_j$ is treated as an ancillary statistic.  We accommodate unobserved subgroups by assigning an arbitrarily large (effectively infinite) standard error and setting the interaction effect to zero (e.g. \(g_j =0\) and \(\sigma_{Bj} = 100\) if subgroup \(B\) is missing), leaving the within-trial contrast well defined. 

In the following, we will consider (subgroup-) \emph{contributions}~$\pi_j$ in a study based on the corresponding \emph{information fraction} [\cite{Spiessens2010}], which is defined based on the standard errors~$\sigma_{Aj}$ and~$\sigma_{Bj}$ as
\begin{equation} \label{eqn:IF}
  \pi_j \;=\; \frac{\sigma_{Bj}^{-2}}{\sigma_{Aj}^{-2} + \sigma_{Bj}^{-2}} \mbox{.}
\end{equation}
This notion may become more intuitive if we consider the common case of a constant \emph{unit information standard deviation} (UISD)~$\uisd$ within a study [\cite{RoeverEtAl2021}]. Standard errors often result (approximately) as $\sigma_{ij}\approx\frac{\uisd}{\sqrt{n_{ij}}}$, where $n_{ij}$~is the number of subjects in study~$j$'s subgroup~$i$. In that case we have
\begin{equation} \label{eqn:IFapprox}
  \pi_j 
  \;\approx\; \frac{n_{Bj} / \uisd^2}{n_{Aj} / \uisd^2 + n_{Bj} / \uisd^2} \;=\; \frac{n_{Bj}}{n_{Aj}+n_{Bj}} \;=\; p_{j} \mbox{,} 
\end{equation}
where $p_j$ simply denotes the \emph{proportion of subjects} in subgroup~$B$. The relationship in  \eqref{eqn:IFapprox} is used only to interpret information fraction as (approximate) prevalence 
%when subgroup counts are available;
when the UISD assumption holds;
CAMS itself is only based on subgroup-specific effect estimates and standard errors 
%(or equivalent summaries) 
and does not necessarily require UISDs. 
%for identification of the interaction.
Instead of subject counts~($n_{ij}$), the UISD expression sometimes may also hold for similar figures, such as \emph{event counts} or \emph{exposures}.
Note that~$\uisd$ here does not need to be the same across trials, but only within trials.
In such cases, subgroup prevalences~$p_{j}$ may serve as a proxy for the contribution~($\pi_j$), although prevalence-based formulations [\cite{Soerensen2023}] do not in general induce the likelihood factorization introduced in the following.

\subsubsection{The problem with varying study contributions}
\begin{figure}
  \centering    
  \includegraphics[width=0.475\textwidth]{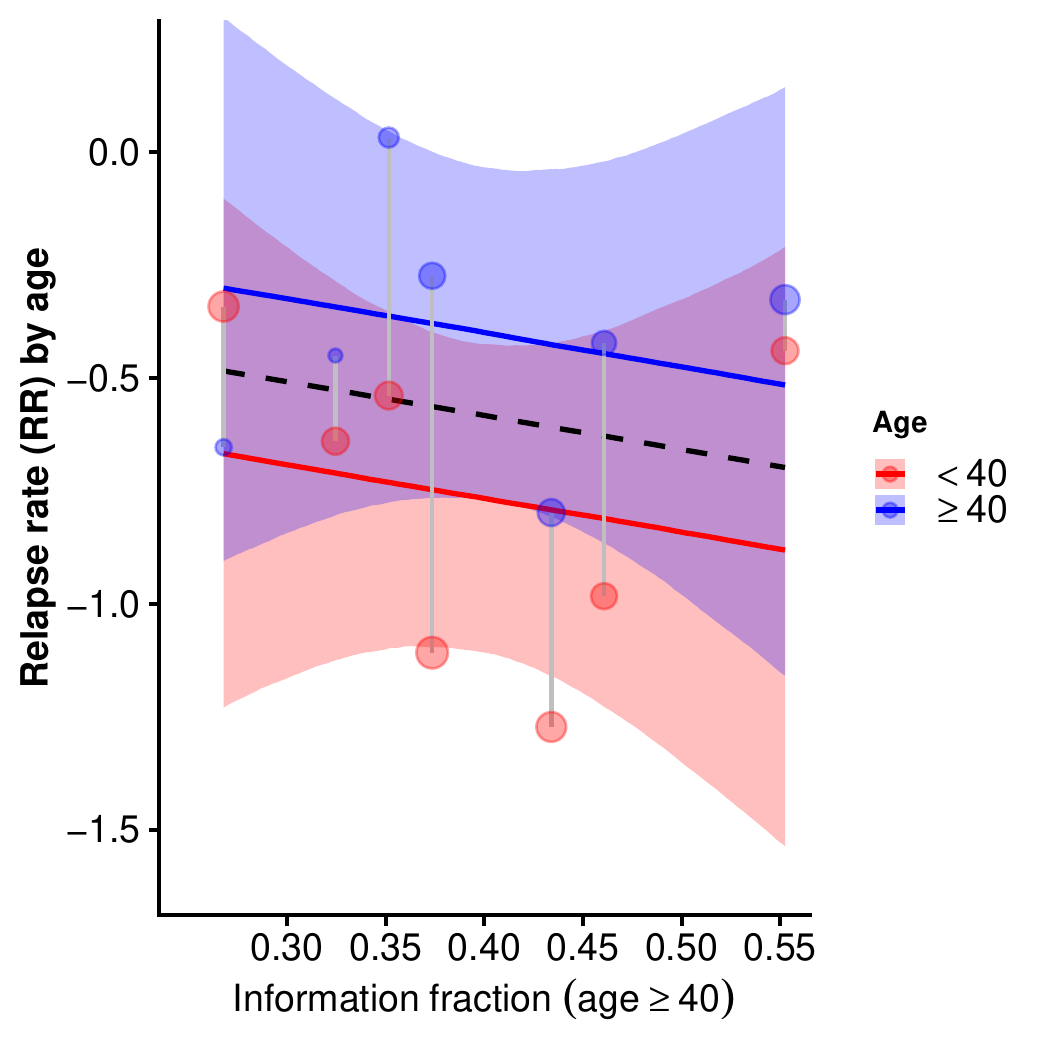}
  \includegraphics[width=0.475\textwidth]{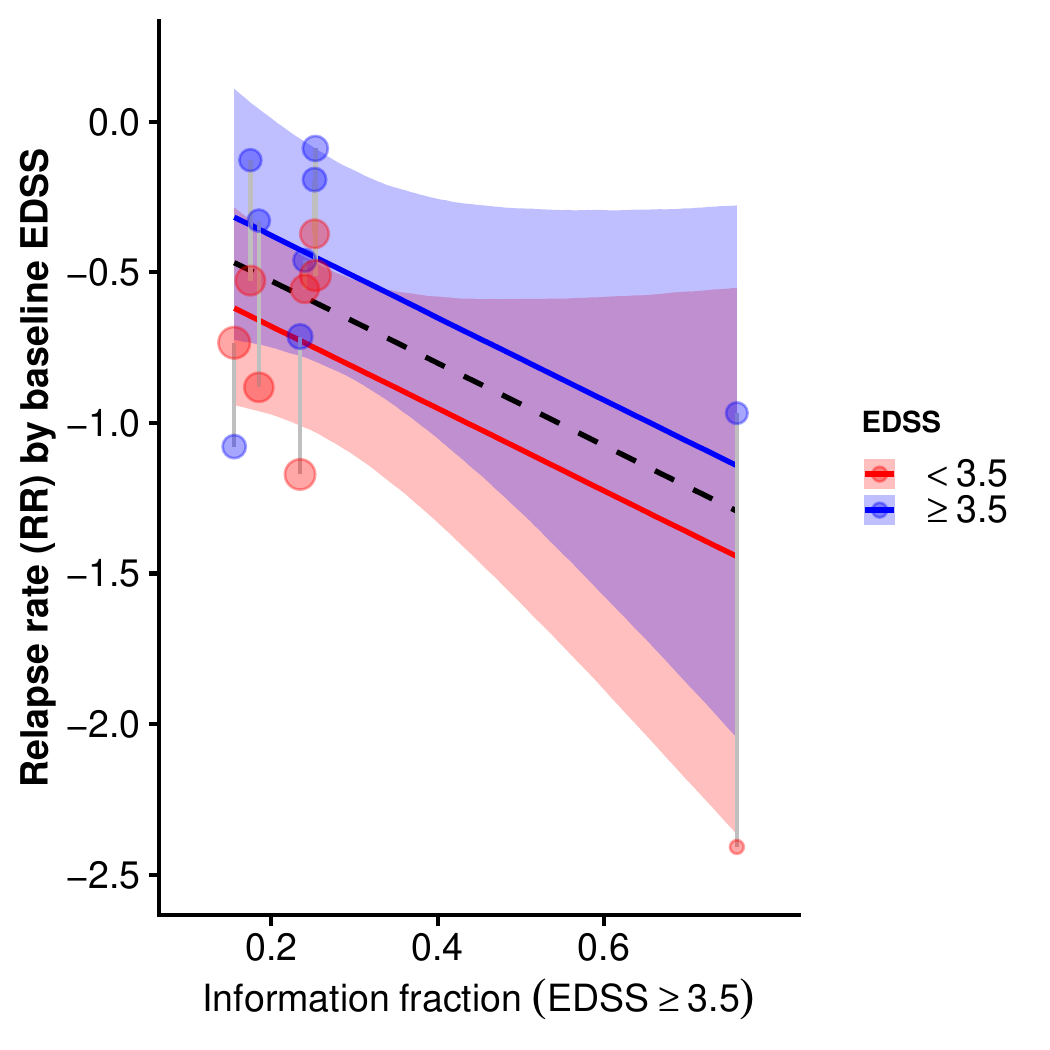}
  \caption{Bubble plots of subgroup-specific effects (conditional treatment effect) on relapse rate (RR) by information fraction. Left: age subgroups (data from Figure~\ref{fig:age-forest}, $\mbox{age}<40$ vs. $\mbox{age}\geq 40$); right: baseline EDSS subgroups (data from Figure~\ref{fig:edss-forest}, $\mbox{EDSS} < 3.5$ vs. $\mbox{EDSS} \ge 3.5$). Coloured lines represent pooled subgroup effects as functions of studies' contributions, with shaded areas showing 95\% pointwise credible intervals. Bubble sizes are proportional to inverse-variance weights, and bubbles from the same trial are connected by (vertical) gray lines, with the vertical distance between them representing the observed treatment-subgroup interaction. The black dashed line indicates the resulting overall effect (marginal treatment effect) . These plots illustrate how subgroup composition can influence the aggregated association, leading to potential aggregation bias in synthesis-generated evidence}
  \label{fig:within_trial_contrasts}
\end{figure}

Figure~\ref{fig:within_trial_contrasts} illustrates why explicitly accounting for how much each study contributes to subgroup comparisons is essential for valid inference about subgroup differences. Each study provides a within-trial contrast (the comparison between subgroups within the same randomized experiment), but studies differ in their contribution: some trials contribute mostly younger patients, others mostly older patients, and contributions also depend on the number of observed events in each subgroup. For example, for endpoints such as mortality, disease recurrence, or time-to-event outcomes like overall survival, subgroups with few events contribute less information to the effect estimate, even if the subgroup sample size is large. The \emph{ecological slope} (dashed line) represents the across-study association between study contribution and study effect estimates: for 0\% or 100\% subgroup contribution, we expect to see the pure subgroup effect, while for a mixed population, we expect something in between [\cite{Robinson1950, Goodman1953, Glynn2010}]. 
Because contribution can correlate with other study-level characteristics (for example, geography, case severity, or trial inclusion criteria), a non-zero ecological slope driven by such confounding signals the presence of aggregation bias: the pooled estimate then conflates the true within-trial subgroup difference with spurious between-study variation. Explicitly decomposing the total association into (i)~the within-trial contrast and (ii)~the ecological (between-study) slope allows us to separate the within-trial subgroup difference (identified by within-trial contrasts) from spurious between-study associations driven by confounding.

The most recent study included in the EDSS meta-analysis (right panel of Figure~\ref{fig:within_trial_contrasts}, and COP1 PILOT in Figure~\ref{fig:edss-forest}), which differs in subgroup prevalence and information fraction from the rest, acts as a leverage point in the bubble plot. While in the forest plot it appears as just another study, in the contribution-adjusted view it exerts disproportionate influence on the overall slope, illustrating how aggregation bias can emerge from varying subgroup contributions.

\subsubsection{Non-collapsibility}
\label{sec:Noncolap}
Within-trial mismatch can arise because of the well-known non-collapsibility problem of some effect measures~[\cite{Greenland2010}]. Subgroup-specific and overall estimates do not necessarily combine in a simple additive way, as represented by \(m_j\) in Section~\ref{sec:Datastructureandnotation}. This may occur, for example, when separate survival models are fitted for each subgroup and for the overall population. When the effect measure is non-collapsible, the relationship
\begin{eqnarray}\label{eqn:overalles}
m_j &=& y_{Aj} + \pi_j g_j
\end{eqnarray}
need not hold exactly, and we treat it as a modeling assumption. Throughout the development of CAMS, we assume \emph{within-trial collapsibility} of the chosen effect measure, so that~\eqref{eqn:overalles} holds within each trial \(j\). 

\subsection{Meta-analysis of Interactions}
\label{sec:Bayesianinteractionmetaanalysisinteraction-MA}
We first consider meta-analysis of the treatment-by-subgroup interactions using the data structure and notation established in Section~\ref{sec:Datastructureandnotation}. We focus on the study-wise interaction estimates and analyze these using standard (univariate random-effects) meta-analysis. Interaction-MA offers a direct estimate of the treatment-by-subgroup interaction effect across studies, with minimal assumptions about the heterogeneity structure. This approach involves conducting a univariate meta-analysis directly on subgroup difference estimates within trials.
    \begin{equation}
         g_{j} = \gamma_j + \epsilon_j, \quad
        \epsilon_j \vert \sigma_{ij}\sim \normaldistn\left(0,~ \sigma_{Aj}^2 + \sigma_{Bj}^2 \right),\quad
        \gamma_j \vert \gamma, \tau_\gamma \sim \normaldistn\left(\gamma,~\tau_\gamma^2\right)
    \end{equation}
so marginally
  \( g_j \vert \sigma_{ij}, \gamma, \tau_\gamma  \sim \normaldistn(
    \gamma,  
    \sigma_{Aj}^2 + \sigma_{Bj}^2 + \tau_\gamma^2)\),
where we denote the trial-specific interaction as \(g_j := y_{Bj} - y_{Aj}\) and we assume that \(\text{Cov}(\epsilon_j, \gamma_j) =0\). We estimate the interaction \(\gamma\) with minimal assumptions.  It cannot alone provide subgroup-specific effects. 

\subsection{Meta-Analysis by Subgroup}
\label{sec:Bayesianmetaanalysisbysubgroupsubgroup-MA}
Having defined a model for meta-analysis of interactions (Section~\ref{sec:Bayesianinteractionmetaanalysisinteraction-MA}), we now turn to a model for meta-analysis of subgroups. To handle subgroup heterogeneity, we adopt the structured random‐effects covariance (``Model 4'') recommended by Jackson et al. (2018), which showed strong performance in simulation studies [\cite{JacksonEtAl2018}],
    \begin{eqnarray}
      && y_{ij}  \;=\; \alpha_j + \gamma_j\bigl(x_{ij}-c_j\bigr)  + \varepsilon_{ij},\quad
      \varepsilon_{ij} \vert \sigma_{ij} \;\sim\; \normaldistn\bigl(0,~ \sigma_{ij}^2\bigr), \nonumber \\
      && \alpha_j \mid \alpha, \tau \;\sim\; \normaldistn\bigl(\alpha,~ \tau^2\bigr), \quad  
      \gamma_j \mid \gamma, \tau_\gamma \;\sim\; \normaldistn\bigl(\gamma,~ \tau_\gamma^2\bigr)  \mbox{,} \label{eqn:subgroup-MA}
    \end{eqnarray}
where the subgroup-specific within-trial variance is denoted as~$\sigma_{ij}^2$, with
$\sigma_{ij}^2=\sigma_{Aj}^2$ when $x_{ij}=0$ (subgroup $A$) and $\sigma_{ij}^2=\sigma_{Bj}^2$ when $x_{ij}=1$ (subgroup $B$). Here $x_{ij}\in\{0,1\}$ indicates subgroup membership, and $c_j$ is the subgroup-centering constant, commonly set to $c_j = 0.5$.
Although originally proposed for arm-based meta‐analysis, its compound-symmetry form carries over naturally to the subgroup setting and facilitates a joint subgroup-MA formulation rather than separate subgroup fits. 

While straightforward and widely used, the pooling of marginal subgroup effects has important drawbacks and can produce mismatches between subgroup and interaction effects [\cite{Berlin2002, RueckerSchumacher2008, Godolphin2022}]. In Figure~\ref{fig:comparison}, for example, the meta-analysis under the subgroup-MA framework yields subgroup (RR) effect estimates of~\(0.62\) for patients with higher disability and~\(0.49\)
for those with lower disability, corresponding to an interaction (RRR) estimate of approximately 
\(1.26\). By comparison, the interaction-MA produces a value of 
\(1.35\), indicating evidence in favour of greater benefit among patients with lower disability\@. 

\subsection{Contribution-adjusted Meta-analysis by Subgroup (CAMS)}
\label{sec:IFadjustedBayesianmeaanalysisbysubgroupCAMS}

The CAMS model avoids the borrowing of information between subgroup-specific means and the interaction effect.  Consequently, the within-trial contrast \(g_j = y_{Bj} - y_{Aj}\) is sufficient for \((\gamma, \tau_\gamma)\) when comparing subgroups by pooling the summary vector \((y_{Aj}, y_{Bj})^\prime\). Instead of enforcing agreement through weighted-average estimators, we use a model-based adjustment. This adjustment reconciles within-trial contrasts with subgroup means and separates ecological trends from within-trial differences.
    
\subsubsection{Implicit ecological slope} \label{sec:Implicitecologicalslope} 
The implicit ecological slope model is obtained by including an ecological slope in the contribution-centred (\(c_j = \pi_j\)) model of~(\ref{eqn:subgroup-MA}). A similar parameterisation, separating within- and across-trial information, has been used in frequentist one-stage IPD modeling~[\cite{FisherEtAl2011, RileyEtAl2020}]. 
\begin{eqnarray}
     && y_{ij} \;=\; \alpha_j + \beta \pi_{j}  + \gamma_j\bigl(x_{ij}-\pi_{j}\bigr)  + \varepsilon_{ij}, \quad  
     \varepsilon_{ij} \mid \sigma_{ij} \;\sim\; \normaldistn\bigl(0,~ \sigma_{ij}^2\bigr), \nonumber \\  
     && \alpha_j \mid \alpha, \tau \;\sim\; \normaldistn\bigl(\alpha,~ \tau^2\bigr), \quad  
     \gamma_j \mid \gamma, \tau_\gamma \;\sim\; \normaldistn\bigl(\gamma,~ \tau_\gamma^2\bigr), \label{eqn:implicitslope}
\end{eqnarray}                   
where $\alpha$~is the baseline effect (e.g.,\ the subgroup~A log-RR), $\gamma$~is the treatment-by-subgroup interaction, and $\beta$~is the  ecological slope across trials and  \(\text{Cov}(\alpha_j, \gamma_j) =0\). 
We can further ensure independent estimation by orthogonalizing the main effect and the interaction.  By orthogonalization we mean projecting the main effect off the interaction so the two estimators are uncorrelated. In practice this is implemented by plugging the information fraction into the design, which yields the required orthogonality under the assumptions stated in Supplementary Section \ref{sec:Equivalenceofinteraction-MAandPrevalenceadjustedsubgroup-MAforinteractioninference}. Alternatively, we define a global slope $\delta(\gamma) = \beta - \gamma$, which we refer to as \emph{explicit slope}.  

\subsubsection{Explicit (global) slope via \(\delta(\gamma)\)}
\label{sec:Explicitslopeviadelta}
 The explicit slope version isolates the within-trial contrast $\gamma_jx_j$ from the entangled component  \(\gamma_j(x_j - \pi_j)\) in~(\ref{eqn:implicitslope}). 
\begin{eqnarray}
       && y_{ij} \;=\; \alpha_j + \delta \pi_{j} + \gamma x_{ij}  + \bigl(\gamma_j - \gamma\bigr)\bigl(x_{ij}-\pi_{j}\bigr)  + \varepsilon_{ij}, \nonumber \quad  
       \varepsilon_{ij} \mid \sigma_{ij} \;\sim\; \normaldistn\bigl(0,~ \sigma_{ij}^2\bigr), \nonumber \\  
       && \alpha_j \mid \alpha, \tau \;\sim\; \normaldistn\bigl(\alpha,~ \tau^2\bigr),  \quad  
       \gamma_j \mid \gamma, \tau_\gamma \;\sim\; \normaldistn\bigl(\gamma,~ \tau_\gamma^2\bigr). \label{eqn:explicitslope}
\end{eqnarray}  
The implicit slope model in~\eqref{eqn:implicitslope} is algebraically equivalent to~\eqref{eqn:explicitslope}, obtained by re\-parametrizing 
the ecological slope as $\delta = \beta - \gamma$. The model nests the unadjusted subgroup-MA when $\beta=0$ or $\pi_j$ is constant (e.g., \(\pi_j = 0.5\)), and reduces to interaction-MA (Section~\ref{sec:Bayesianinteractionmetaanalysisinteraction-MA}) for the marginal distribution of contrasts $y_{Bj}-y_{Aj}$ when $\alpha$ and $\delta$ are treated as nuisance parameters.  In practice, we estimate $\delta(\gamma)=\beta-\gamma$ as an ordinary regression coefficient. Thus, CAMS inherits the unconfounded interaction inference from interaction-MA while retaining the subgroup interpretability of subgroup-MA, resolving the tension between subgroup- and contrast-based approaches. The corresponding marginal distribution of subgroup outcomes is
\begin{eqnarray}
\label{eqn:subgroup_adj}
    \begin{pmatrix}
        y_{Aj}\\
        y_{Bj}\\
    \end{pmatrix}
    \mid d_{ij}  
    &\sim& \normaldistn\left(
    \begin{pmatrix}
    \alpha + (\beta-\gamma)\,\pi_j \\[4pt] 
    \alpha + (\beta-\gamma)\,\pi_j + \gamma    
    \end{pmatrix}, ~ 
    \begin{pmatrix}
    \sigma_{Aj}^2 & 0\\
    0 & \sigma_{Bj}^2
    \end{pmatrix} + \text{T}+ \text{T}_{\gamma,j} \right),
\end{eqnarray}
with $d_{ij} = (\sigma_{ij}, \alpha, \beta, \gamma, \pi_j, \tau, \tau_\gamma)^\prime$, and where the overall and interaction effect heterogeneities are given respectively as
\begin{equation}
  T  \;=\; \begin{pmatrix}\label{eqn:hetmatrix}
    \tau^2 & \tau^2\\
    \tau^2 & \tau^2
    \end{pmatrix}    
  \quad \text{and}\quad 
  T_{\gamma,j}  \;=\;     \begin{pmatrix}
    \pi_j^2~\tau_\gamma^2 & - \pi_j(1 - \pi_j)~\tau_\gamma^2\\
    - \pi_j(1 - \pi_j)~\tau_\gamma^2 & (1-\pi_j)^2~\tau_\gamma^2
    \end{pmatrix}.    
\end{equation}  
Among the CAMS versions, the explicit-slope parametrization \eqref{eqn:explicitslope} is written in a cell-mean, ANOVA-like form, which makes the interaction structure more transparent and is convenient for software specification (e.g., using 
%\texttt{lme4} [\cite{PinheiroBates}]–style 
formula syntax in the \texttt{brms} \textsf{R}~package [\cite{Buerkner2017}]).  Table~\ref{tab:methods-comparison} summarizes the main distinctions among interaction-MA, subgroup-MA and CAMS\@. 

\begin{table}
\centering
\caption{Comparison of three approaches to subgroup meta-analysis. ``Aggregation (ecological) bias'' refers to bias that can arise when across-trial differences in subgroup prevalence contribute information about subgroup differences, thereby mixing within-trial interaction with between-trial case-mix differences}
\label{tab:methods-comparison}
\footnotesize
\begin{tabular}{p{2cm}p{3.5cm}p{3.5cm}p{3.5cm}}
\toprule
Method & Data required & Target estimands & Description \\
\midrule
Interaction-MA
& Within-trial interaction estimates (e.g., log-RRRs) with SEs \emph{or} sufficient subgroup-specific information to form the within-trial contrast
& Interaction $(\gamma,\tau_\gamma)$ only
& Uses within-trial contrasts by construction; does not yield pooled subgroup-specific effects \\

Subgroup-MA
& Subgroup-specific treatment effects (e.g., log-RRs) with SEs (and preferably subgroup counts/prevalences)
& Subgroup effects $(\mu_A, \mu_B)$; interaction as $\mu_B-\mu_A$
& Interaction is purely within-trial only for constant information fraction across trials; otherwise within- and between-trial information can be {conflated} \\

CAMS
& Subgroup-specific treatment effects with SEs (only SEs are needed to derive the information fraction)
& Subgroup effects and interaction $(\gamma,\tau_\gamma)$ jointly
& Targets within-trial interaction by adjusting for subgroup contributions, reducing leakage from between-trial information differences; overall and heterogeneity are independently estimated \\

\bottomrule
\end{tabular}
\end{table}

\subsection{Reporting of pooled subgroup effects} \label{sec:Pooledsubgroupeffectsreporting}
\subsubsection{Information fraction dependence of estimates}

In contribution–adjusted subgroup-MA, subgroup effects depend on the information fraction, whereas the interaction parameter~$\gamma$ remains constant (see also Figure~\ref{fig:within_trial_contrasts}).
Consideration of this contribution-dependence is therefore required for inference on subgroup and overall effects. Assuming a certain \emph{reporting information fraction}~$\pi_{\star}$, the corresponding subgroup-specific effects are
\begin{eqnarray}
\mu_A(\pi_{\star}) = \alpha + \delta\,\pi_{\star},
\quad \mbox{and} \quad
\mu_B(\pi_{\star}) = \alpha + \delta\,\pi_{\star}+ \gamma \mbox{,}
\end{eqnarray}
and the overall (study-level) effect is the contribution-weighted subgroup average
\begin{eqnarray}\label{eqn:overalleffects}
m(\pi_{\star}) = (1-\pi_{\star})\,\mu_A(\pi_{\star}) + \pi_{\star}\,\mu_B(\pi_{\star}) = \alpha + (\delta + \gamma)\pi_{\star} \mbox{,}
\end{eqnarray}
while the treatment-by-subgroup interaction remains constant ($\mu_B(\pi_{\star}) - \mu_A(\pi_{\star}) = \gamma$).

Broadly, the choice may prioritize efficiency, robustness to trial composition, alignment with prior literature, or other external considerations. Ideally, however, the choice should be principled enough to rely on a definition that can yield values that are representative of the expected population prevalence (see Table \ref{tab:flexprev}). We consider three ways of handling dependence on the information fraction (or its proxy, the prevalence) \(\pi\):
(i) report inference at a single value of \(\pi\) that is considered relevant;
(ii) average inference with respect to an assumed or estimated distribution \(p(\pi_j)\) for the proxy prevalences (e.g., from external evidence or a previous meta-analysis), thereby propagating uncertainty in \(\pi\); or
(iii) choose \(\pi\) such that the overall effect estimate is recovered (which requires a within-trial collapsibility assumption). As a simple sensitivity analysis within the empirically supported range, one may additionally report $\mu_A(\pi)$ at $\pi_{\min}=\min_j \pi_j$ and $\pi_{\max}=\max_j \pi_j$ (avoiding extrapolation beyond the observed contributions).

\subsubsection{Overall effect estimate}
\label{sec:Closenesstotheoveralleffectestimate}\label{sec:Closenesstotheoveralleffectestimate}

By \eqref{eqn:sufficiency}, the information fraction separates the interaction terms from the overall effects. Hence, under the proposed parameterisation, the overall effects in \eqref{eqn:overalleffects} are orthogonal to the interactions: their estimates are uncorrelated (and independent under normality), so the overall effect and the interaction carry separate information and can be estimated without one distorting the other. Hence, there exists an intermediate~information fraction~\(\overallif\) at which the factorized model reproduces the standard univariate overall-effect meta-analysis based on inverse-variance weighting of the ancillary marginals~\(m_j\). 
This corresponding \emph{overall information fraction} is given by
\begin{equation}\label{eqn:overall_IF}
\overallif \;=\; \frac{\sum_j w_j \pi_j}{\sum_j w_j}\mbox{,} 
\quad \mbox{where} \quad
w_j \;=\; \frac{1}{\tau^2 + (1-\pi_j)^2\sigma_{Aj}^2+\pi_j^2\sigma_{Bj}^2} \;\approx\; \frac{n_j}{\tau^2 n_j + \uisd^2} 
\mbox{,}
\end{equation}
which under the assumption of a common UISD~\eqref{eqn:IFapprox} and common-effect model ($\tau=0$) simplifies to the overall prevalence among all studies included in meta-analysis (\(\overallif = \sum_jn_{Bj}/\sum_jn_{j}\)). Here we assume within-trial collapsibility as in~\eqref{eqn:overalles}, and  variances consistent with such decomposition.

Reporting subgroup effects at the overall information fraction has therefore two advantages: it preserves the orthogonality of subgroup contrasts and interactions, and it ensures that reported overall effects remain coherent with univariate meta-analysis. For practical reporting, we therefore recommend subgroup effects to be presented under this information fraction, since it uniquely bridges the subgroup-adjusted and overall analyses while maintaining confounding-free inference on interaction. Reporting the overall information fraction \(\overallif\) alongside the corresponding estimates, and possibly supplementing these with an additional figure, may further improve interpretability.

\subsubsection{Fixed information fraction} \label{sec:Fixedinfo}
\begin{figure}
  \centering
  \includegraphics[width = 0.495\textwidth, trim=0 5mm 0 0mm,clip]{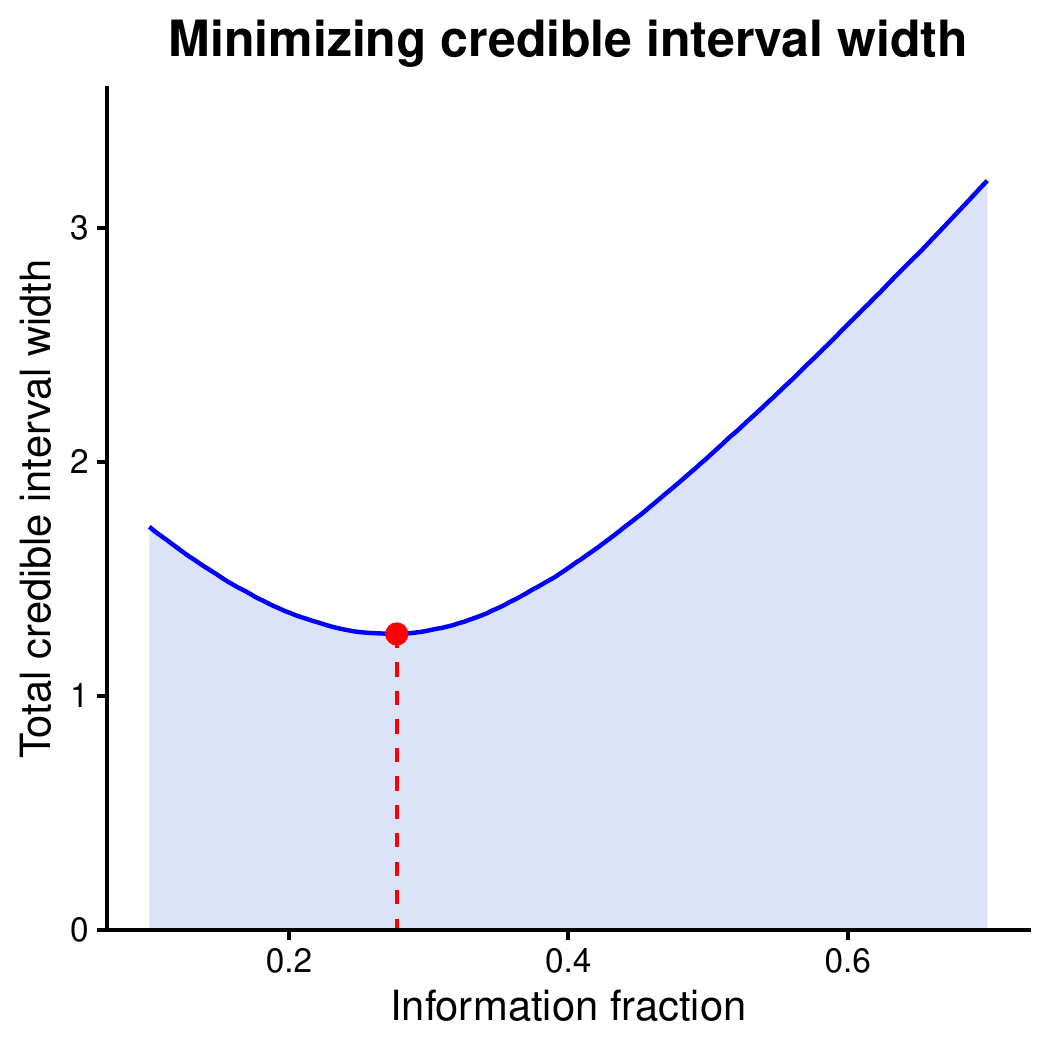}
  \includegraphics[width = 0.495\textwidth, trim=0 5mm 0 0mm,clip]{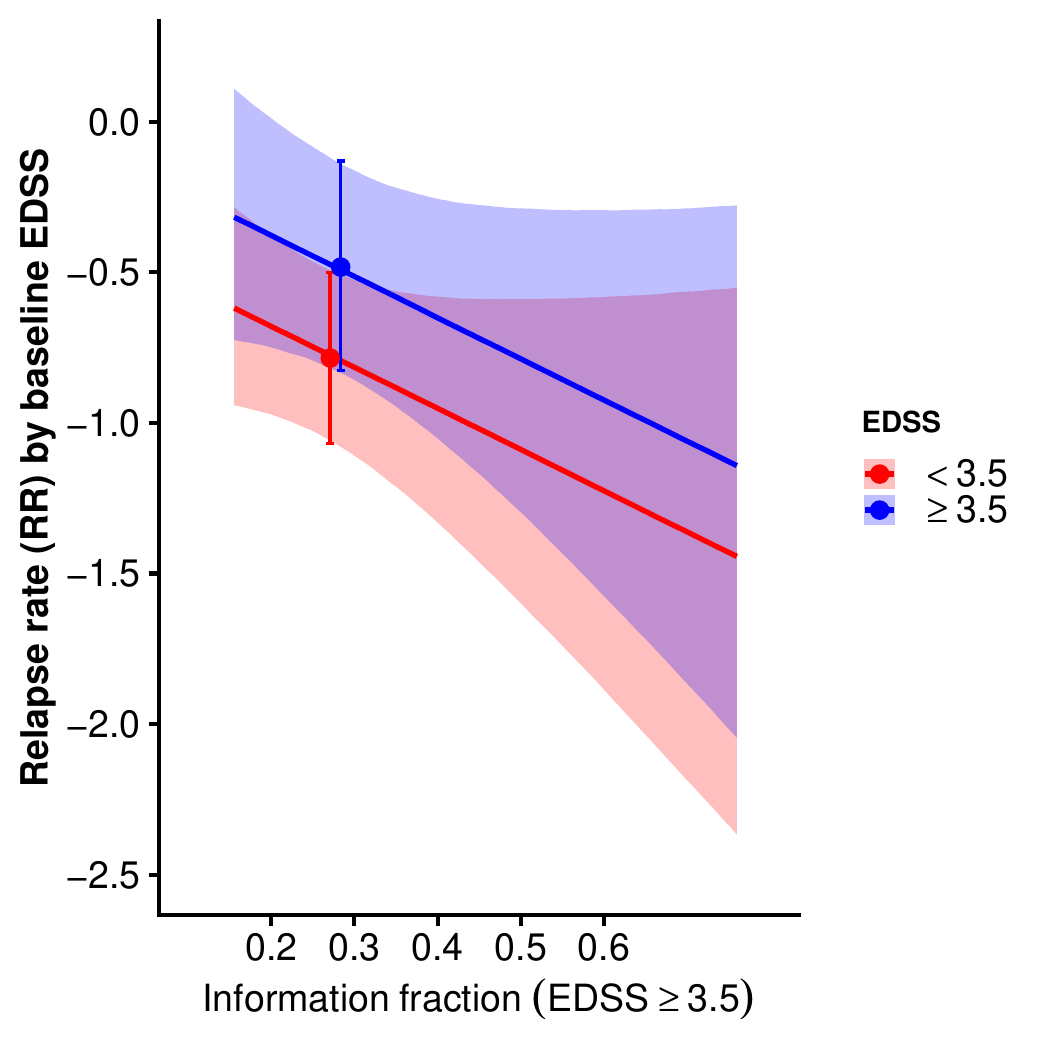}
  \caption{Optimization of the subgroup-specific credible interval (CrI) width.
  \emph{Left}: total credible interval width versus prevalence; the dot marks the minimal-width (optimal) prevalence.
  \emph{Right}: subgroup posterior estimates (EDSS~< 3.5 vs \(\ge\) 3.5) across reporting information fractions with 95\% credible intervals}
  \label{fig:optimal-if}
\end{figure}

The choice of a single reporting~information fraction~$\reportingif$ can be motivated pragmatically.
A natural default may be to use a weighted average of the study-specific contributions as the population prevalence for \(m_j\), which may again coincide with~$\overallif$ \eqref{eqn:overall_IF} in some cases.
%Choosing $\pi=\pi_\star$ ensures that the reported effect corresponds to an information fraction that is strongly supported by the data (when no external information is incorporated).
Reporting subgroup effects at a typical information fraction yields the relatively narrow credible intervals that might also lead to underestimation of uncertainty.
This behaviour is illustrated in
Figure~\ref{fig:optimal-if}
by comparing the posterior distributions of the overall effect under varying information fractions.
The shortest attainable interval may yield an overly optimistic representation of both the overall and the subgroup-specific effects, in particular when information fractions may differ in relevant applications.
Consideration of a range of information fraction configurations can enhance robustness in settings where study composition exerts substantial influence,
which might be relevant if studies considered in the analysis generally do not reflect information fractions relevant in practice, or if the analysis was dominated by single (or few) studies with extreme information fractions, as exemplified by a recent meta-analysis including a large single-subgroup trial~[\cite{Sterne2020}].

Several aspects may be relevant in guiding the choice of a reporting information fraction:
\begin{enumerate}
   \item \emph{Closeness to subgroup estimates} (\(
   \reportingif = \arg\min_\pi \, \vert \mu_{A}(\pi) - \widehat\alpha_{\textrm{subgroup-MA}} \vert
   \)):  
   Select the prevalence $\reportingif$ that minimizes the discrepancy between model-based subgroup summaries and observed subgroup data, where we use the distance between posterior medians. The same construction applies to subgroup~$B$ by relabelling ($A \leftrightarrow B$ and $\pi \leftrightarrow 1-\pi$). This is a clear alternative to the overall-information fraction used in Figure~\ref{fig:comparison}. 

   \item \emph{Plain average prevalence} (\(
   \reportingif = \frac{1}{k} \sum_{j=1}^k \pi_j
   \)):  
   Take the arithmetic mean of subgroup prevalences across trials. This is a straightforward definition of the reporting information fraction, but it may be sensitive to outlying prevalences, especially in single-subgroup trials.

   \item \emph{Trial-weighted prevalence} (   \(
   \reportingif = \sum_{j=1}^k w_j \pi_j/\sum_{j=1}^k w_j
   \)):  
   Compute a weighted mean of subgroup prevalences with weights proportional to inverse-variance interaction weights,  \(
   w_j = \frac{1}{1/n_{Aj} + 1/n_{Bj}}\). The weight $w_j$ is proportional to the effective sample size for a two-subgroup contrast (harmonic mean of $n_{Aj}$ and $n_{Bj}$), and therefore down-weights highly imbalanced or single-subgroup trials. This provides a second alternative to the overall-information fraction, and in particular yields a prevalence that is effectively trial-size weighted when subgroup sizes are similar.

   \item \emph{Technical knowledge-based prevalence}:
%   (\(
%   \reportingif = \pi^{\text{external}}
%   \)):  
   Use external information, e.g., registry or hospital data. This option allows the inclusion of external empirical evidence. For example, in the context of the MS~example, Pugliatti et al.~[\cite{Pugliatti2006}] found that a substantial majority, about 68\%, of patients remain in the low-disability range for a prolonged period.  
\end{enumerate}

\subsubsection{Uncertain subgroup prevalence}
\label{sec:Subgroupprevalencebasedonexternaldata}

Instead of conditioning on a \emph{fixed} information fraction~$\reportingif$, inference may also be based on a probability distribution, in order to reflect uncertainty in~$\reportingif$. 
Given a prevalence distribution~$p(\pi)$ that may be obtained via meta-analysis of internal and/or external evidence, or based on other considerations of plausible $\reportingif$~values, the subgroup effects' joint posterior distribution then results as a \emph{mixture distribution} [\cite{Lindsay}].
We may then for example compute the expectations
\begin{eqnarray}\label{eqn:subgroupreport}
\bar{\mu}_A \;=\; \int \mu_A(\pi) \, p(\pi) \, \mathrm{d}\pi, 
\qquad
\bar{\mu}_B \;=\; \bar{\mu}_A + \gamma.
\end{eqnarray}
This uncertainty propagation is straightforward when using Monte Carlo approximations. First, samples of $(\alpha,\delta,\gamma)^{\prime}$ from the joint posterior of \eqref{eqn:subgroup_adj} may be drawn using routines available e.g. in the \texttt{brms} package; second, a sample of  $\reportingif$ from $p(\pi)$ is generated (e.g., using the \texttt{RBesT} package).
Finally, credible intervals for $(\bar{\mu}_A,\bar{\mu}_B)$ may be computed based on empirical quantiles [\cite{MCMCinPractice}].

In the following results, we used the \texttt{RBesT} package [\cite{RBesT}] to construct a predictive distribution~$p(\pi)$ from internal evidence on subgroup counts; this construction is based on prevalence rather than information fraction. This is achieved through meta-analytic predictive (MAP) synthesis, combining binomial likelihoods with half-normal priors on between-study heterogeneity [\cite{RoeverEtAl2021}]. The MAP distribution provides a coherent prediction for the prevalence, which can then be approximated by a Beta or Beta mixture distribution. \\

The modeling assumptions underlying CAMS, introduced across the preceding subsections, are summarised in Table~\ref{tab:assumptions}.\\

\begin{table}
\centering
\caption{Summary of the main modelling assumptions underlying CAMS, with the section in which each is introduced and its role. Assumptions are stated for the chosen effect measure on which the analysis is performed (e.g., log-rate-ratio, log-odds-ratio, log-hazard-ratio)}
\label{tab:assumptions}
\footnotesize
\begin{tabular}{p{2.5cm}p{1.5cm}p{8cm}}
\toprule
Assumption & Introduced & Role / when it matters \\
\midrule
Normal approximation with known within-trial variances
& Sections~\ref{sec:Bayesianinteractionmetaanalysisinteraction-MA},~\ref{sec:Bayesianmetaanalysisbysubgroupsubgroup-MA} and~\ref{sec:IFadjustedBayesianmeaanalysisbysubgroupCAMS}
& Subgroup estimates $y_{Aj}, y_{Bj}$ are approximately normal on the chosen effect scale (e.g., log-OR, log-RR, mean difference), and the standard errors $\sigma_{Aj},\sigma_{Bj}$ are treated as fixed and known, as in standard meta-analysis. Enables the likelihood in \eqref{eqn:subgroup-MA}. \\

Handling of unobserved subgroups
& Section~\ref{sec:Datastructureandnotation}
& A trial reporting only one subgroup is accommodated by assigning an effectively infinite standard error (and zero interaction contribution) to the missing subgroup, leaving the within-trial contrast well defined. Similar assumption is made in Godolphin et al.~[\cite{Godolphin2022}].\\

Information fraction as the contribution/centring quantity
& Section~\ref{sec:IFadjustedBayesianmeaanalysisbysubgroupCAMS}
& $\pi_j$ is defined from the standard errors as the within-trial information fraction~[\cite{Spiessens2010}], and the subgroup indicator is centred on $\pi_j$ rather than on the covariate mean as in one-stage IPD models~[\cite{RileyEtAl2020}]. Under a constant UISD~[\cite{RoeverEtAl2021}] it reduces to subgroup prevalence, but CAMS requires only the standard errors. \\

Within-trial collapsibility of the chosen effect measure
& Section~\ref{sec:Noncolap}
& The overall effect equals the contribution-weighted combination of subgroup effects, $m_j = y_{Aj} + \pi_j g_j$. Automatic for collapsible measures (e.g., mean difference, risk difference, risk ratio) but a non-trivial working assumption for non-collapsible measures ~[\cite{Greenland2010}], notably the odds ratio and hazard ratio; matters mainly for the overall-effect and reporting steps. \\

Independence of overall and interaction random effects
& Section~\ref{sec:Implicitecologicalslope}, Appendix~\ref{sec:Equivalenceofinteraction-MAandPrevalenceadjustedsubgroup-MAforinteractioninference}
& The subgroup random effects follow the ``Model~4'' structure of Jackson et al.~[\cite{JacksonEtAl2018}], giving one overall heterogeneity $\tau$ and one interaction heterogeneity $\tau_\gamma$. $\mathrm{Cov}(\alpha_j,\gamma_j)=0$, and priors are placed independently on the overall and interaction blocks. Coupling them (or fixing $\pi_j=0.5$) breaks the factorization and the agreement with interaction-MA. \\

\bottomrule
\end{tabular}
\end{table}

\section{Application to example data}
\label{sec:Results}

\subsection{Relapse rate reduction by age and disability score subgroups}
\label{sec:Resultsofprevalenceadjustment:AgeanddisabilitysubgroupsintheMSmeta-analysis}
\begin{figure}
    \centering
     \includegraphics[width = \textwidth]{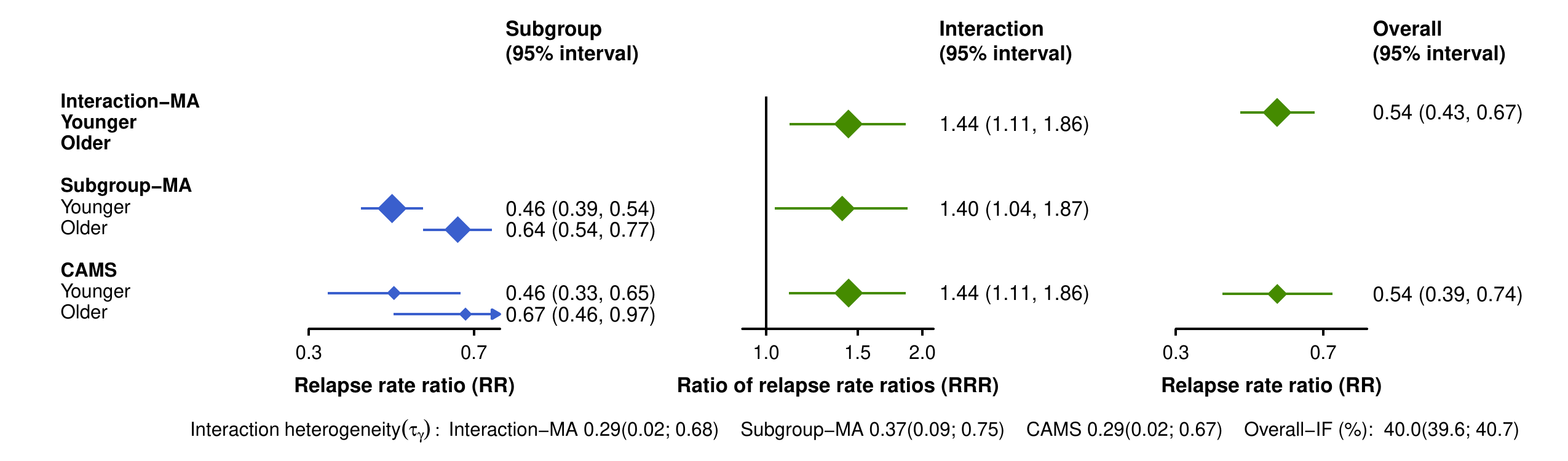}
    \includegraphics[width = \textwidth]{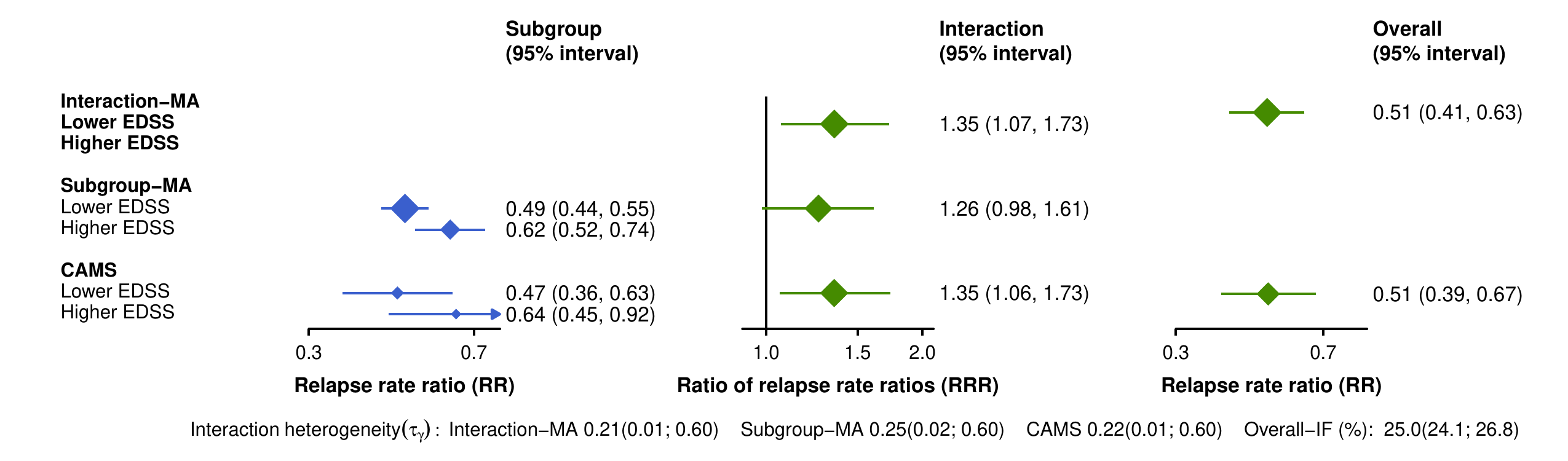}
  \caption{Comparison  between interaction-MA, subgroup-MA, and CAMS estimates for the data sets shown in Figures~\ref{fig:age-forest} and \ref{fig:edss-forest}. Each row corresponds to one method; the upper and lower blocks show relapse rate ratios (RR) by age and EDSS, respectively.  The left panels show pooled subgroup estimates for each  meta-analysis method. The middle panels display pooled interaction effects. The right panels display pooled overall effects. The intervals include reported posterior median with 95\% equal-tailed credible intervals. The priors on heterogeneity parameters were \(\tau \sim \halfnormaldistn(1.0)\) and \(\tau_\gamma \sim \halfnormaldistn(0.5)\). The overall-information fraction is used as discussed in Section \ref{sec:Closenesstotheoveralleffectestimate} }
    \label{fig:comparison}
\end{figure}

Figure~\ref{fig:within_trial_contrasts} shows that one small study exerts strong leverage on the ecological slope estimate in the EDSS example, whereas no similarly outlying study appears in the age-based analysis.
As previously noted, the interaction estimates in Figure~\ref{fig:comparison} differ between methods (subgroup-MA vs interaction-MA). For consistency with the original analysis and to facilitate direct comparison with the previously reported results, the analyses and figures presented in the main text retain the Bayesian formulations of interaction-MA and subgroup-MA. Corresponding frequentist analyses are provided in Supplementary Appendix. 

In the therapy-by-age meta-analysis, subgroup-MA yields an interaction of $\mbox{RRR}=1.40$,
while Interaction-MA and CAMS give slightly larger estimates of $\mbox{RRR}=1.44$.
Similarly, in the meta-analysis by disease severity (EDSS), the subgroup-MA estimates the interaction as $\mbox{RRR}=1.26$ (with 1.0 included in the credible interval), whereas Interaction-MA and CAMS yield a larger interaction ($\mbox{RRR}=1.35$) as well as credible intervals excluding an RRR of~$1.0$.
Section~\ref{sec:Noncolap} discusses this mismatch; the difference between the age and EDSS datasets mainly reflects inclusion of a small EDSS study with an outlying subgroup contribution for the high-EDSS group (subgroup prevalence~$48\%$, information fraction~$70\%$). The relationship between CAMS and the frequentist within-trial framework of Godolphin et al.~[\cite{Godolphin2022}], with which it may coincide in the common-effect limit, is discussed in the Conclusion (Section ~\ref{sec:Conclusion}).

The two meta-analyses use essentially the same studies (EDSS adds only one small study), so the overall effects should be almost identical regardless of subgrouping variable. This is confirmed in Figure~\ref{fig:comparison}, where overall estimates are about 0.49–0.54 for both age and EDSS subgrouping. However, the overall results are not numerically identical: the small decimal differences between the age- and EDSS-based overall estimates reflect the impact of the within-trial collapsibility assumption. Its influence on the overall effect is negligible because the study is small, but it can still be influential in the interaction analysis when its subgroup contribution is atypical.

\begin{figure}[!h]
  \centering
  \includegraphics[width = .33\textwidth]{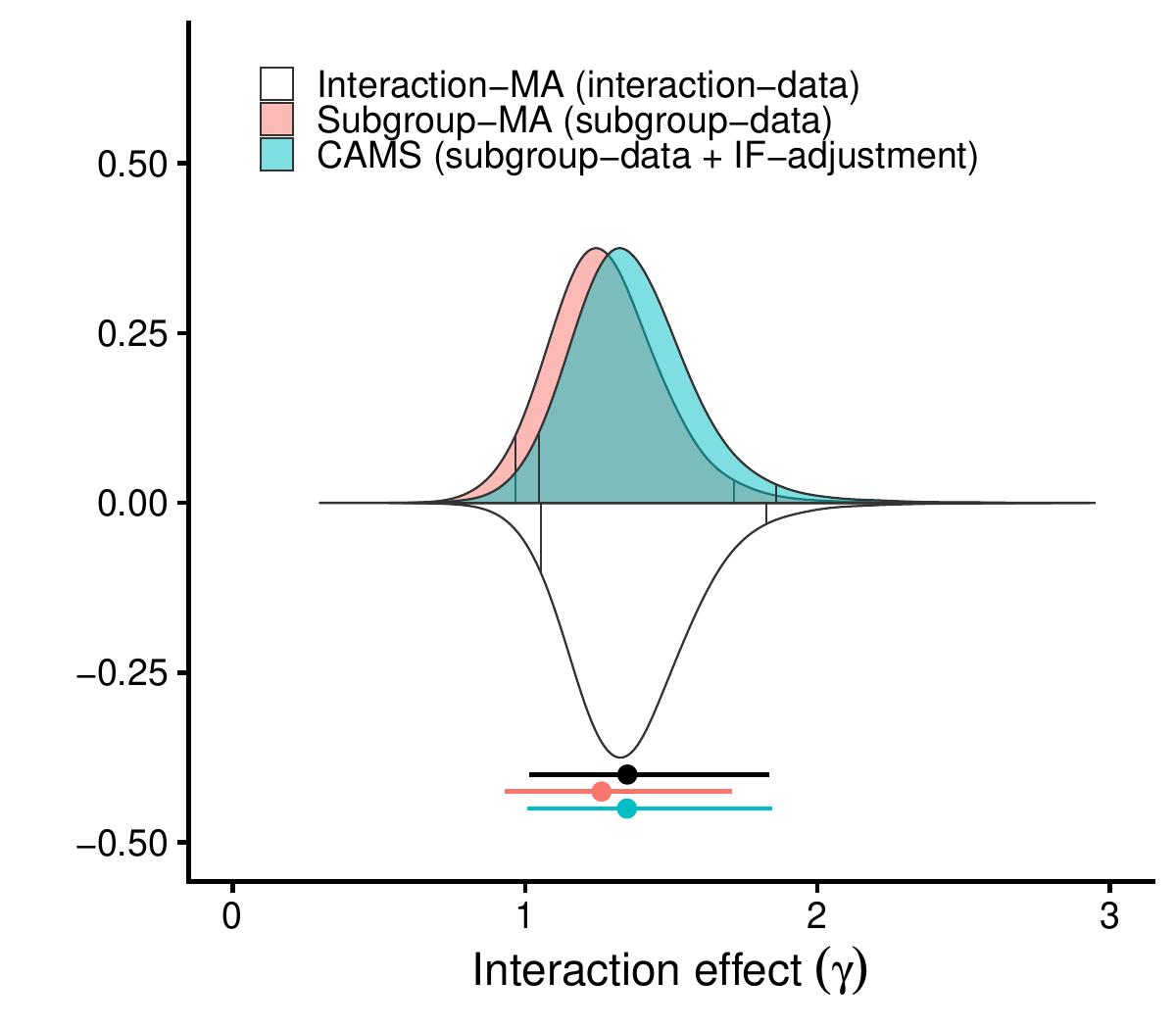}
  \includegraphics[width = .33\textwidth]{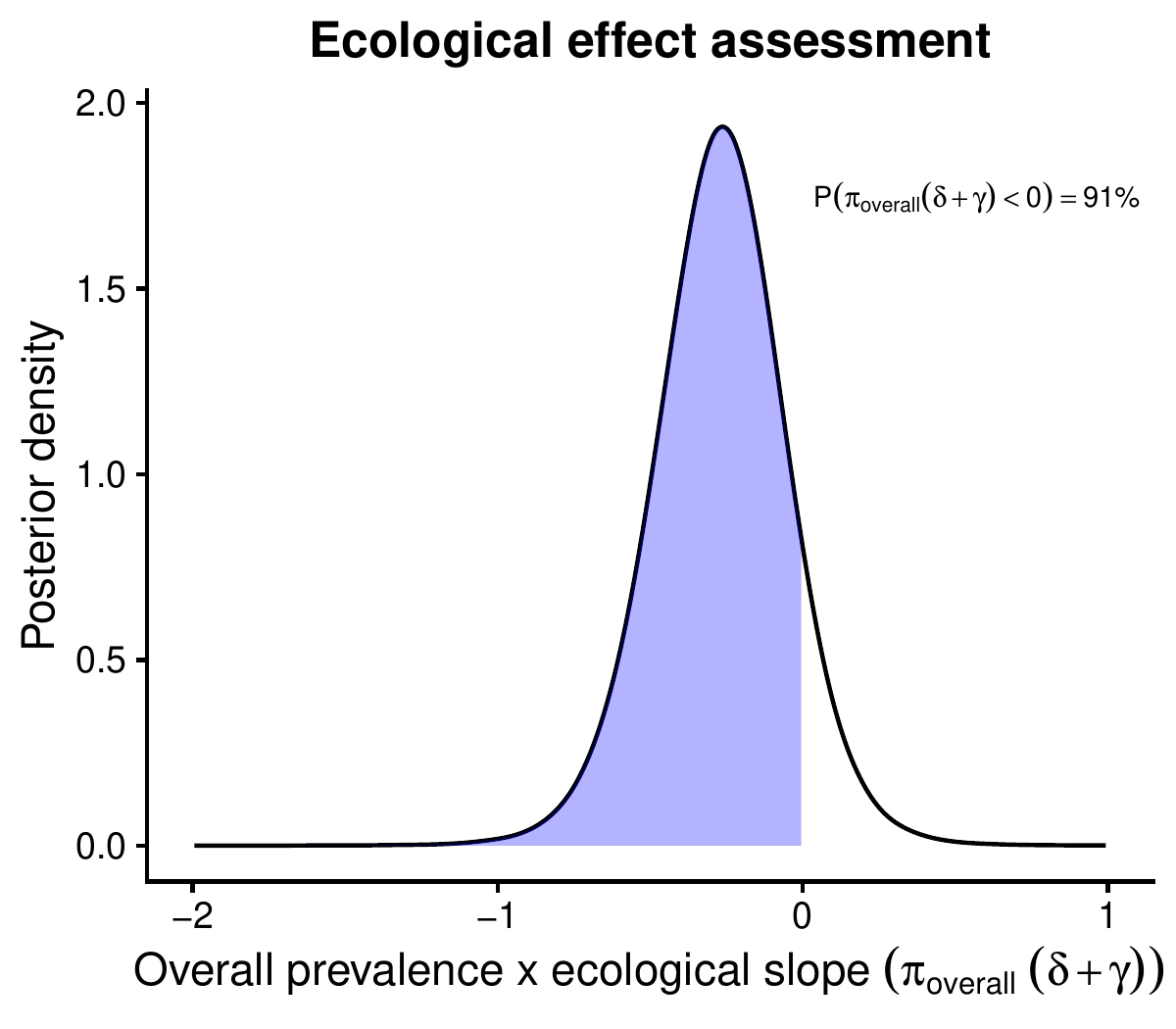}
  \includegraphics[width = .33\textwidth]{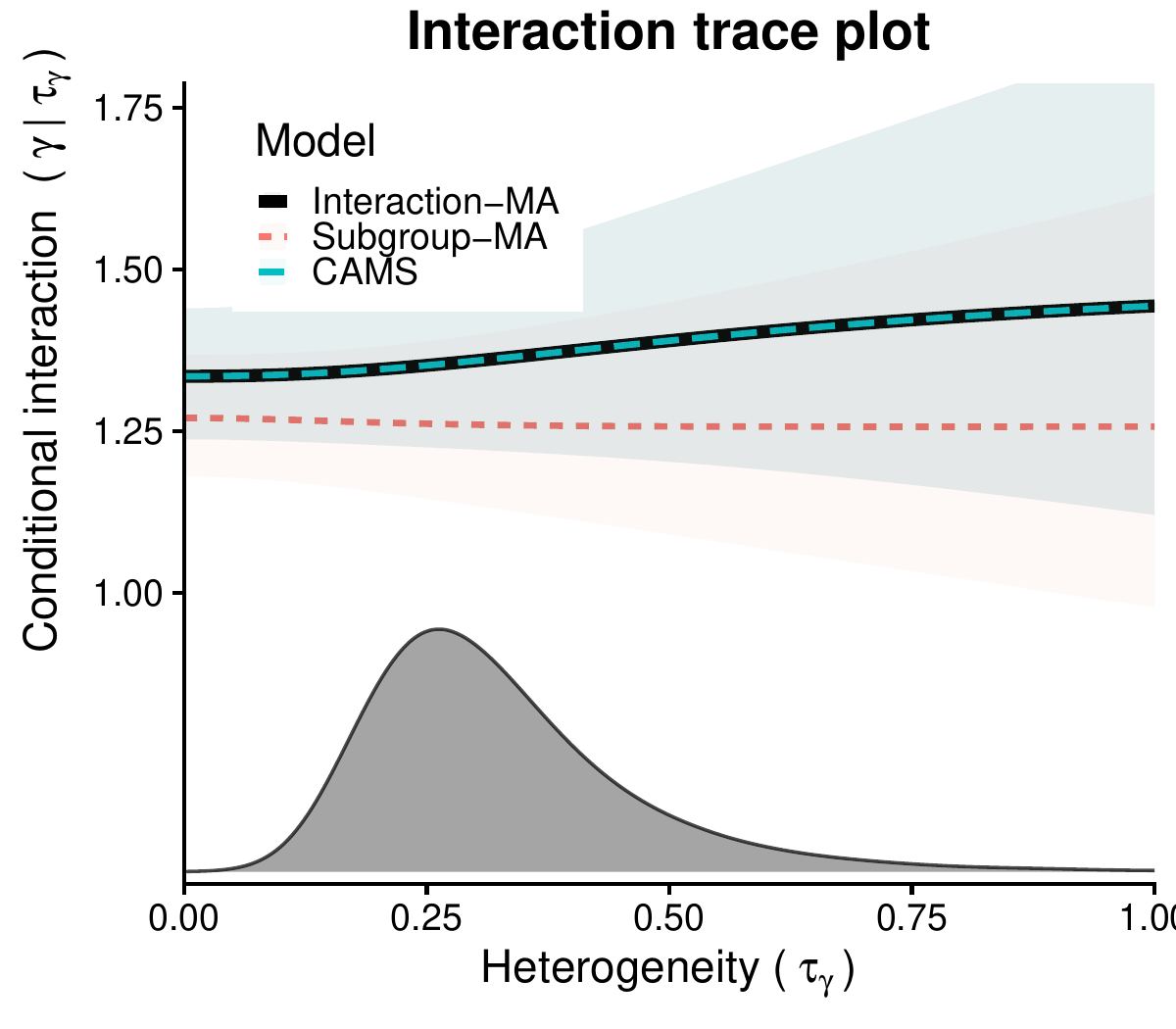}
  \caption{Posterior results. \emph{Left:} Posterior densities compared, subgroup-MA on subgroup-data can recover  interaction-MA inference when contribution-adjusted, that is the CAMS (contribution-adjusted subgroup-MA). 
  \emph{Middle:} Tail probability of an ecological association (evidence of aggregation bias at the study level), \(91\%\) in the EDSS analysis (Figure~\ref{fig:edss-forest}). 
  \emph{Right:} Conditional posterior distribution of the interaction effect with shaded 50\% pointwise credible intervals}
  \label{fig:aggregation-bias}
\end{figure}

The ecological slope introduced in Section~\ref{sec:Explicitslopeviadelta} can capture spurious correlations between outcomes and subgroup contributions. It may therefore be used not only as a diagnostic but also as a comparative device. We examine tail posterior probabilities in both examples (Figures~\ref{fig:age-forest} and \ref{fig:edss-forest}), thereby quantifying the impact of including the small study. Figure~\ref{fig:aggregation-bias} displays posterior summaries for the EDSS meta-analysis (small study present) under the elicited overall-information fraction from Figure~\ref{fig:comparison}. The middle panel shows a posterior ecological effect (evidence of aggregation bias) of 91\% in the EDSS analysis (Figure~\ref{fig:edss-forest}) compared to a corresponding value of 31\% for the age meta-analysis (small study absent).

Turning to the right panel, the conditional distribution of the interaction effect is shown as a function of its heterogeneity \(\tau_\gamma\). The trace plot~[\cite{RoeverRindskopfFriede2024}] shows that for any given interaction heterogeneity \(\tau_\gamma\), both interaction-MA and CAMS exhibit overlapping (cyan and black) trajectories while the unadjusted subgroup-MA (red) fails to do so. Agreement between interaction-MA and CAMS also extends to heterogeneity estimates (Figure~\ref{fig:comparison}) and is supported by the factorization in \eqref{eqn:sufficiency}, provided the same prior distributions are specified. These figures illustrate agreement in the analyses shown and show that the proposed contribution adjustment can reconcile subgroup-based, contrast-based, and overall approaches within a unified likelihood framework. These results are consistent with the equivalence in \eqref{eqn:sufficiency} under matched priors, and illustrate agreement between interaction-MA and CAMS across the heterogeneity range examined here. 

\subsection{Sensitivity analyses}
\label{sec:Sensitivityanalysis}

\begin{figure}
\centering
\includegraphics[width=\textwidth]{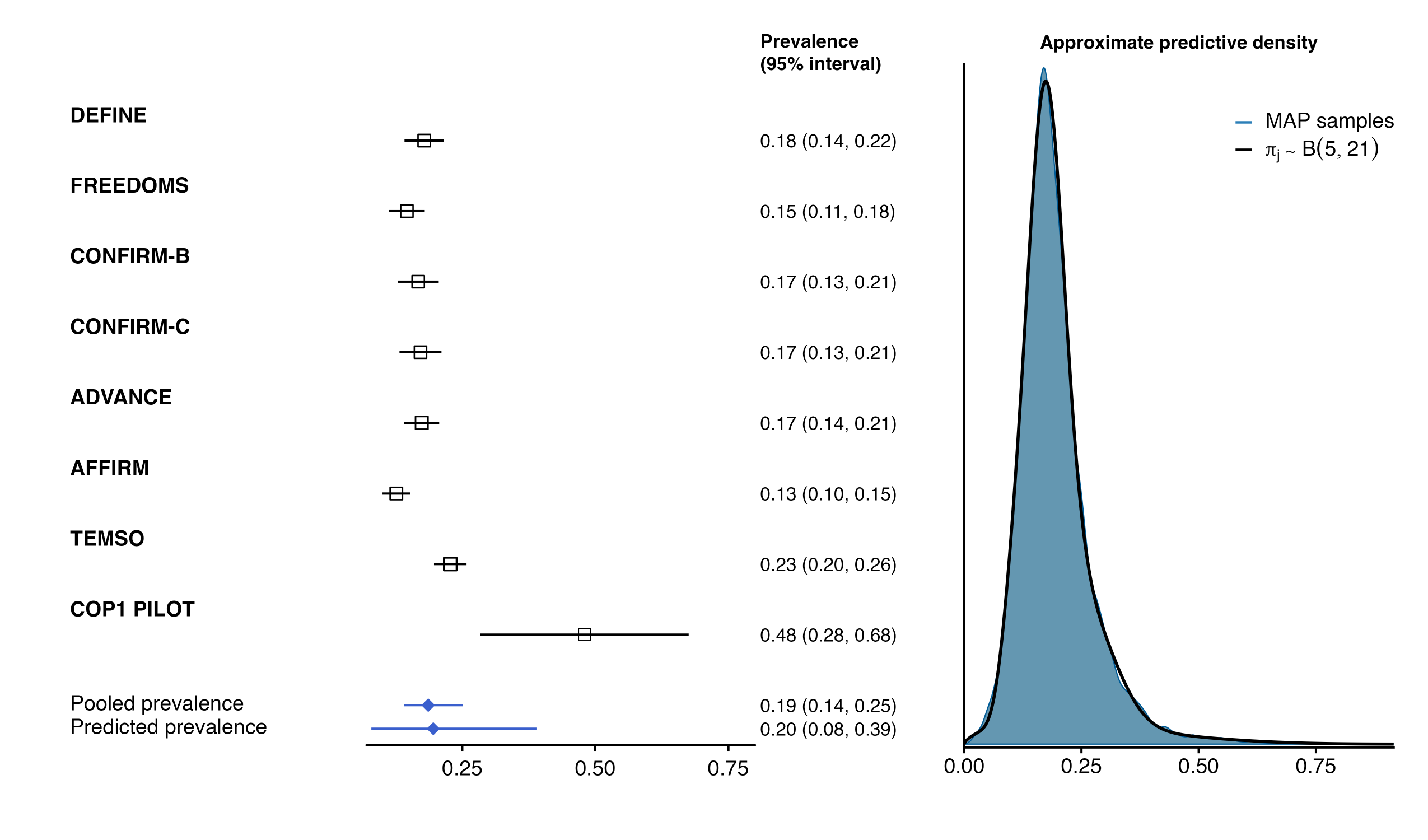}
\caption{Study-specific subgroup prevalences as information fraction surrogate with 95\% confidence intervals (black boxes and lines). 
The blue diamond and bar denote the pooled prevalence, while the long blue interval shows the predictive prevalence for a new trial. The overlaid $\mathrm{Beta}(5,21)$ distribution is used for reporting, mean of \(19.2\%\) and standard deviation of \(7.6\%\). It has less concentrated mass on central values but lighter tails providing robustness against outlying prevalence values. The horizontal axis represents proxy information fraction (subgroup prevalence) (0--1)}
\label{fig:map}
\end{figure}

We assess sensitivity to the choice of the reporting information fraction~$\reportingif$ based both on a fixed number or a distributional  assumption. Table~\ref{tab:flexprev} reports subgroup-specific estimates under these different prevalence definitions. For choices that propagate uncertainty (Section~\ref{sec:Subgroupprevalencebasedonexternaldata}), the reported information fraction has an associated standard deviation: an overall-IF of 25.0\% with standard deviation 0.72\%, from~\eqref{eqn:overall_IF}; and a MAP prevalence of 19.2\% with standard deviation 7.6\%, from the Beta(5,21) distribution of Figure~\ref{fig:map}. Figure~\ref{fig:optimal-if} illustrates determination of an optimal (optimistic) information fraction (second line Table \ref{tab:flexprev}), as described in Section~\ref{sec:Fixedinfo}. The left panel shows the total credible interval width as a function of information fraction, reaching its minimum at around $27.7\%$ for the high-disability subgroup. The right panel shows the relapse rate estimates by disability (EDSS) subgroup, previously shown in Figure~\ref{fig:within_trial_contrasts}, and indicates where the 95\% subgroup credible interval would be defined under this prevalence specification. This value differs from the overall-information fraction of $\overallif = 25.
0\%$ adopted for reporting in Figure~\ref{fig:comparison}.

Across methods, subgroup estimates remain broadly consistent, although small shifts are visible across prevalence choices. 
Notably, the interaction effect estimate remains constant, while low- and high-EDSS estimates vary slightly. 
This illustrates that prevalence specification affects subgroup-level inferences, even in cases where the overall conclusions remain unchanged.  The reporting information fraction is an explicit assumption and should be stated transparently when presenting subgroup-specific pooled effects. This is illustrated by the analysis based on the reported overall prevalence in Figure~\ref{fig:comparison}. 
%The next section introduces a distributional perspective on subgroup prevalence, providing an additional principled framework for subgroup estimate reporting based on a MAP prevalence distribution.

While the shortest interval may offer an ``optimal'', but also optimistic choice in terms of posterior precision, it entails important caveats. In single-subgroup trials where $\pi_j \in \{0,\,1\}$ (i.e., only one subgroup is represented), the estimate of the optimal information fraction can be distorted by the absence of the other subgroup, since the entire weight is placed on a single subgroup and the optimal prevalence may therefore be driven by an extreme (near-boundary) unobserved composition. If reported outside the observed range, inference on $\reportingif$ effectively amounts to extrapolation beyond observed trial compositions. Conceptually, the optimum corresponds to the prevalence that is best supported by \emph{the amount of available data} regardless of its quality, in contrast to the \emph{overall-information fraction}, which follows a trial-weighted direction.

Figure~\ref{fig:map} (left panel) shows trial-specific subgroup prevalences with 95\% intervals for the high-EDSS subgroups (see also Figure~\ref{fig:edss-forest}). 
There is substantial between-trial variation (0.13–0.48), with COP1~PILOT showing a notably higher prevalence than the remaining studies. The pooled prevalence (blue diamond) is centred around 0.19 with a 95\% interval of 0.14--0.25, whereas the predictive prevalence for a new trial (long blue interval) is much wider, 0.20 (0.08--0.39). The right panel visualizes this predictive distribution for prevalence; the overlaid Beta(5,21)~density provides reasonable approximation to the predictive distribution that we utilize in the following [\cite{RBesT}].

When the CAMS posterior is averaged over the predictive prevalence distribution (MAP prevalence in Table \ref{tab:flexprev}), the point summaries 
change a bit (closer to the nearest, trial-weighted $\reportingif=0.22$), and uncertainty in~\(\pi\) is propagated through to the subgroup effect estimates. Consequently, effect estimates (overall and for both subgroups) have wider intervals, while the interaction estimate (as expected) remains the same. 

\begin{table}
\centering 
\caption{Comparison of subgroup-specific relapse rate estimates under alternative \emph{post-hoc} prevalence elicitation. 
Each approach specifies the distribution of low- versus high disability differently (efficient intervals, closeness to subgroup estimates, arithmetic mean, trial-weighted mean, external knowledge, or clinical expertise). Posterior medians with 95\% (equal-tailed) credible intervals are reported, showing that different prevalence choices lead to small but measurable differences in subgroup effect estimates while interaction is preserved by design, see Figure~\ref{fig:optimal-if}}
\footnotesize\begin{tabular}{p{2.25cm} p{1.5cm} p{2cm} p{2cm} p{2cm} p{2cm}}
\toprule
{Method} & {Information fraction (IF)} & {Low-EDSS} & {High-EDSS} & Overall & Interaction \\ 
\midrule
Overall-IF & 25.0\% & 0.47 (0.35; 0.62) & 0.64 (0.45; 0.91) & 0.51 (0.39; 0.66) & 1.35 (1.07; 1.73) \\
Optimal/optimistic IF & 27.7\% & 0.46 (0.34; 0.61) & 0.62 (0.44; 0.88) & 0.50 (0.38; 0.65) & 1.35 (1.07; 1.73) \\
Closer to Subgroups & 25.0\% & 0.47 (0.36; 0.63) & 0.64 (0.45; 0.92) & 0.51 (0.39; 0.67) & 1.35 (1.07; 1.73) \\
Average (naive) & 28.3\% & 0.45 (0.34; 0.60) & 0.61 (0.43; 0.87) & 0.49 (0.38; 0.65) & 1.35 (1.07; 1.73) \\
Trial-weighted & 22.0\% & 0.49 (0.37; 0.66) & 0.67 (0.46; 0.97) & 0.53 (0.40; 0.70) & 1.35 (1.07; 1.73) \\
Clinical Expertise & 31.9\% & 0.43 (0.32; 0.58) & 0.58 (0.41; 0.83) & 0.47 (0.36; 0.63) & 1.35 (1.07; 1.73) \\
MAP prevalence & 19.2\% & 0.51 (0.35; 0.77) & 0.69 (0.45; 1.12) & 0.54 (0.38; 0.80) & 1.35 (1.07; 1.73) \\
 \bottomrule
\end{tabular}
\label{tab:flexprev}
\end{table}

\section{Conclusions}
\label{sec:Conclusion}

Estimates from the contribution-adjusted meta-analyses by subgroup (CAMS) provide interaction-MA equivalent interaction inference and coherent subgroup estimates, and therefore offer a practical alternative method. In practice, CAMS combines the main strengths of the two standard approaches: it retains robust interaction inference from interaction-MA, while yielding interpretable subgroup summaries that align with the estimated interaction.

The CAMS framework may also be applied within a frequentist framework, but the number of studies included would need to be sufficiently large for reliable estimation of the variance components. In particular, random-effects meta-analyses based on only two or three studies are especially challenging, because between-study heterogeneity is then difficult to estimate~[\cite{KnappHartung2003, RoeverKnappFriede2015}]. In a frequentist analysis this is typically addressed by the Hartung--Knapp adjustment, which accounts for the uncertainty in the estimated heterogeneity; we instead regard the Bayesian formulation, with priors on the heterogeneities and full uncertainty propagation, as particularly attractive in this few-study setting. When many studies are available, the heterogeneities \(\tau\) and \(\tau_\gamma\) can be estimated reliably, for example by DerSimonian--Laird or restricted maximum likelihood~[\cite{DerSimonianLaird1986, Viechtbauer2005}] applied to \(m_j\) and \(g_j\). A simpler alternative in either case is to plug the variances in as fixed values, or to set them to zero in the common-effect limit, at the cost of not propagating their uncertainty.

The contribution-adjusted parameterisation also has a close connection to one-stage IPD models that separate within- and across-trial information~[\cite{FisherEtAl2011, Hua2017, RileyEtAl2020}]. In those models, centring is based on the study-specific covariate mean, corresponding to subgroup prevalence for a binary covariate, whereas CAMS uses the within-trial information fraction. This distinction is important because the information-fraction centring yields the factorisation that preserves interaction and interaction-heterogeneity inference while retaining subgroup-specific effects. CAMS adapts this separation to the aggregate-data setting, where related approaches have also sought to avoid aggregation bias in subgroup inference [\cite{Godolphin2022, Soerensen2023}]. Under the within-trial collapsibility assumption, it also provides an overall information fraction at which the conventional univariate overall-effect meta-analysis is recovered. The proposed framework is also not restricted to random-effects models and may be used in common-effect formulations when appropriate~[\cite{SchmidCarlinWelton2021}]. In this common-effect limit, the estimates may coincide with the ones obtained with the framework of Godolphin et al.~[\cite{Godolphin2022}]. However, the two approaches differ in how they report subgroup-specific effects. CAMS expresses these effects as an explicit function of the reporting information fraction, allowing the contribution to be chosen so that it matches the target population. In contrast, the within-trial framework does not provide this choice. This distinction is important when the observed subgroup prevalences are not representative of the target population, for example when some trials report only one subgroup, or when the observed composition leads to a misleading contribution, as in the EDSS example in Figure \ref{fig:map}. In such cases, CAMS can report subgroup effects at a population-relevant contribution rather than one driven by the observed trial compositions. Common effects may be implemented either for subgroup effects or interactions (or both). For both frequentist and Bayesian analyses, routines are straightforward with readily available standard \textsf{R}~packages, for example with \texttt{metafor} [\cite{Viechtbauer2010}] and \texttt{brms} [\cite{Buerkner2017}]. Extensions to multi-arm [\cite{BretzEtAl2005}] and network settings [\cite{DiasAdes2016}] are promising directions for future research. In addition, further work on prior specification may be valuable, particularly in settings with only a small number of trials, where prior assumptions can have greater influence [\cite{TurnerEtAl2012, RoeverEtAl2021}].

\appendix

\clearpage
\begin{appendix}
\section{Appendix}

\subsection{Equivalence of interaction-MA and CAMS for interaction inference}
\label{sec:Equivalenceofinteraction-MAandPrevalenceadjustedsubgroup-MAforinteractioninference}
We now show that incorporating study contributions into the analysis guarantees orthogonality between contrasts and means, and the CAMS likelihood factorizes such that interaction inference coincides with interaction-MA\@. A key requirement of any interaction meta-analysis is that conclusions about the interaction should not depend on whether we first model subgroup means and then take their difference (subgroup-MA), or directly model the contrast between subgroups (interaction-MA).

In the CAMS formulation, the study contribution~$\pi_j$ is chosen such that interaction and overall effects $(g_j,m_j)$ are orthogonal.  This allows for both frequentist and Bayesian inference, only requiring that independent priors are placed on subgroup-specific and interaction parameters (without mixing across-trial and within-trial borrowing). This construction guarantees that CAMS recovers interaction-MA for~$\gamma$. If, instead, $\pi_j$~is fixed at~$\pi_j = 0.5$ (or if the priors couple the subgroup and interaction blocks), the factorization no longer holds and subgroup-MA ceases to coincide with interaction-MA.

We show that, for a unique choice of study-level prevalence~$\pi_j$, CAMS coincides with interaction-MA in both likelihood and posterior inference for the interaction parameters $\gamma$ and $\tau_\gamma$. Define the sufficient within-study contrast and ancillary mean [\cite{Basu1977}]  as
\begin{eqnarray}
g_j = y_{Bj} - y_{Aj}, \qquad 
m_j = (1-\pi_j)\,y_{Aj} + \pi_j\,y_{Bj}.
\end{eqnarray}
By the Neyman-Fisher factorization [\cite{Neyman1935}], the joint density factorizes as
\begin{eqnarray}
  p(g_j,m_j \mid \alpha, \beta, \gamma, \pi_j, \tau, \tau_\gamma)
  &=& p(g_j \mid \gamma,\tau_\gamma)\;p(m_j \mid \alpha,  \beta, \tau, \pi_j),
\end{eqnarray}
which implies that $g_j$ is a sufficient statistic for $(\gamma,\tau_\gamma)$. Thus, all information about $(\gamma,\tau_\gamma)$ is contained in the one-dimensional contrast $g_j$. Because the linear transformation from $(y_{Aj},y_{Bj})$ to $(g_j,m_j)$ is nonsingular, no information is lost [\cite{Jackson2013o}]. This construction mirrors the use of orthogonal contrasts in classical factorial Analysis of Variance (ANOVA) [\cite{Tjur1984}], where main effects and interactions are represented by mutually orthogonal linear combinations of cell means so that each effect is estimated from its own independent component of the data.

In our setting, \(g_j\)~plays the role of a \emph{within-trial} interaction contrast, while \(m_j\)~corresponds to a prevalence-weighted main-effect averaged \emph{across-trials}; use of the information fraction ensures that these two are orthogonal, and hence that inference on~\((\gamma, \tau_\gamma)\) is free from cross-study information borrowing. Under CAMS, this factorization leads to
\begin{eqnarray} \label{eqn:sufficiency}
  \begin{pmatrix}g_j\\m_j\end{pmatrix}
  \,\Big|\, d_{ij} \sim
  \normaldistn\left (
   \begin{pmatrix}\gamma\\[3pt]\alpha + \beta\,\pi_j\end{pmatrix},~
    \begin{pmatrix}
      \tau_\gamma^2 + \sigma_{Aj}^2 + \sigma_{Bj}^2 & 0\\[6pt]
      0 & \tau^2 + (1-\pi_j)^2 \sigma_{Aj}^2 + \pi_j^2\sigma_{Bj}^2
    \end{pmatrix}
  \right ),
\end{eqnarray}
which can be extended by introducing a heterogeneity parameter on the overall effect.
Orthogonality (and, under normality, independence) requires that $\mathrm{Cov}(g_j,m_j)=0$, which holds if and only if the information fraction is used,
\begin{eqnarray}\label{eqn:uisd}
  \pi_j = \frac{\sigma_{Aj}^2}{\sigma_{Aj}^2 + \sigma_{Bj}^2} \mbox{,}
%  \approx \frac{\uisd^2/n_{Aj}}{\uisd^2/n_{Aj} + \uisd^2/n_{Bj}} = \frac{n_{Bj}}{n_{Bj} + n_{Aj}}.    
\end{eqnarray}
where \(\pi_j\)~approximates the subgroup prevalence (of subgroup~$B$) in trial~$j$ as in \eqref{eqn:IFapprox}. 
Rather than in a generic regression parametrization, the interaction specification is driven by orthogonal contrasts of within-trial subgroup effects, while subgroup and overall means are collected in a separate, orthogonal block. For any other specification (arbitrary~$\pi_j$), one can show that
\begin{eqnarray}
\operatorname{Cov}(g_j,m_j)
\;=\; \pi_j \sigma_{Bj}^2 - (1-\pi_j)\sigma_{Aj}^2 \quad \mbox{if~} \textstyle \pi_j \neq \frac{\sigma_{Aj}^2}{\sigma_{Aj}^2 + \sigma_{Bj}^2} \mbox{.}
\end{eqnarray}

Orthogonality of~$g_j$ and~$m_j$ is therefore achieved if and only if the information fraction is used as the (adjusted) ecological slope covariate. 
In the special case 
of~\(c=0.5\), \eqref{eqn:subgroup_adj}~reduces to~\eqref{eqn:subgroup-MA}, and the covariance is given by
\begin{eqnarray}\label{eqn:fixed}
\mathrm{Cov}(g_j,m_j)
\;=\; 
%[\,1,\,-1\,]\;\Sigma_j\;\begin{pmatrix}0.5\\0.5\end{pmatrix} =
\tfrac12~\bigl(\sigma_{Aj}^2 - \sigma_{Bj}^2\bigr)\mbox{.}
\end{eqnarray}
Hence, the factorization also holds for a subgroup-MA only when all trials are \emph{perfectly balanced}, i.e.\ $\sigma_{Aj}^2=\sigma_{Bj}^2$ (and hence \(\pi_j=0.5\)) .  
From \eqref{eqn:sufficiency}, it is worth noting that interaction estimation depends exclusively on within-trial differences~$g_j$, while the overall trial summaries~\(m_j\) are sufficient for the overall effect. Thus, inference matching is guaranteed not only for the interaction effects, which were our primary focus, but also for the overall effects (and their respective heterogeneities, if modelled), provided that within-trial collapsibility is assumed (see Section \ref{sec:Noncolap}). 

\subsection{Frequentist analyses}

For completeness, frequentist versions of interaction-MA, subgroup-MA and CAMS are reported in
Table~\ref{tab:frequentist}, together with results from the within-trial framework~[\cite{Godolphin2022}]. The Bayesian versions are retained in the main text to maintain correspondence with the original analysis in Heesen et al.~[\cite{Heesen2024}].

\begin{table}[ht]
\centering
\caption{Frequentist analyses for the age and EDSS subgroup comparisons.
Estimates are relapse-rate ratios (RR) for subgroups and the overall effect, and ratios of RRs (RRR) for the interaction, with 95\% Wald confidence intervals.
Between-study heterogeneity was estimated by restricted maximum likelihood (REML).
The overall effect is available only from CAMS, evaluated at each information fraction (IF).
Subgroup A is age ${<}40$ or EDSS ${<}3.5$}
\label{tab:frequentist}
\footnotesize\begin{tabular}{llcccc}
\hline
Subgroup  & Analysis & Subgroup A & Subgroup B & Overall & Interaction \\
\hline
Age 
 & Interaction-MA & -- & -- & -- & 1.44 (1.09; 1.90) \\
 & Subgroup-MA & 0.46 (0.39; 0.55) & 0.64 (0.53; 0.78) & -- & 1.40 (1.02; 1.92) \\
 & CAMS (Overall-IF 39.6\%) & 0.47 (0.37; 0.59) & 0.67 (0.51; 0.88) & 0.54 (0.43; 0.67) & 1.44 (1.09; 1.90) \\
 & CAMS (Optimal IF 39.7\%) & 0.47 (0.37; 0.59) & 0.67 (0.51; 0.88) & 0.54 (0.43; 0.67) & 1.44 (1.09; 1.90) \\
 & CAMS (Closer to Subgroups 43.3\%) & 0.45 (0.35; 0.59) & 0.65 (0.49; 0.87) & 0.53 (0.42; 0.67) & 1.44 (1.09; 1.90) \\
 & CAMS (Average (naive) 39.4\%) & 0.47 (0.37; 0.59) & 0.67 (0.51; 0.88) & 0.54 (0.43; 0.67) & 1.44 (1.09; 1.90) \\
 & CAMS (Trial-weighted 40.9\%) & 0.46 (0.36; 0.59) & 0.66 (0.51; 0.87) & 0.54 (0.43; 0.67) & 1.44 (1.09; 1.90) \\
 & CAMS (Clinical Expertise 31.9\%) & 0.49 (0.37; 0.67) & 0.71 (0.50; 1.00) & 0.55 (0.42; 0.74) & 1.44 (1.09; 1.90) \\
 & CAMS (MAP prevalence 19.2\%) & 0.54 (0.31; 0.94) & 0.78 (0.43; 1.41) & 0.58 (0.34; 1.00) & 1.44 (1.09; 1.90) \\
 & Within-trial framework & 0.46 (0.35; 0.61) & 0.67 (0.55; 0.81) & -- & 1.44 (1.09; 1.90) \\
\hline
EDSS
 & Interaction-MA & -- & -- & -- & 1.34 (1.05; 1.70) \\
 & Subgroup-MA & 0.49 (0.43; 0.56) & 0.62 (0.50; 0.76) & -- & 1.26 (0.96; 1.66) \\
 & CAMS (Overall-IF 25.3\%) & 0.47 (0.38; 0.58) & 0.63 (0.48; 0.83) & 0.51 (0.42; 0.62) & 1.34 (1.05; 1.70) \\
 & CAMS (Optimal IF 25.9\%) & 0.47 (0.38; 0.58) & 0.63 (0.48; 0.82) & 0.51 (0.41; 0.62) & 1.34 (1.05; 1.70) \\
 & CAMS (Closer to Subgroups 24.7\%) & 0.48 (0.39; 0.59) & 0.64 (0.49; 0.84) & 0.51 (0.42; 0.63) & 1.34 (1.05; 1.70) \\
 & CAMS (Average (naive) 28.3\%) & 0.45 (0.36; 0.56) & 0.61 (0.46; 0.80) & 0.49 (0.40; 0.61) & 1.34 (1.05; 1.70) \\
 & CAMS (Trial-weighted 22.0\%) & 0.49 (0.40; 0.61) & 0.66 (0.50; 0.88) & 0.53 (0.43; 0.65) & 1.34 (1.05; 1.70) \\
 & CAMS (Clinical Expertise 31.9\%) & 0.43 (0.34; 0.55) & 0.58 (0.44; 0.76) & 0.47 (0.38; 0.59) & 1.34 (1.05; 1.70) \\
 & CAMS (MAP prevalence 19.2\%) & 0.51 (0.41; 0.64) & 0.69 (0.51; 0.92) & 0.54 (0.44; 0.68) & 1.34 (1.05; 1.70) \\
 & Within-trial framework & 0.47 (0.36; 0.60) & 0.63 (0.47; 0.83) & -- & 1.34 (1.05; 1.70) \\
\hline
\end{tabular}
\end{table}
\end{appendix}

\clearpage

\begin{Backmatter}
%\paragraph{Acknowledgments}
%We are grateful for the technical assistance of A. Author.

\paragraph{Funding Statement}
  Support from the \emph{Volkswagen Stiftung} is gratefully acknowledged (project ``Bayesian and nonparametric statistics - Teaming up two opposing theories for the benefit of prognostic studies in COVID-19'').

\paragraph{Competing Interests}
  The authors have declared no conflict of interest.

\paragraph{Data Availability Statement}
The data and code supporting the findings of this study are available at \emph{GitHub} under the URL \url{https://github.com/rvpanaro/CAMS}; the code also includes a frequentist implementation of the CAMS approach. 

%\paragraph{Ethical Standards}
%The research meets all ethical guidelines, including adherence to the legal requirements of the study country.

\paragraph{Author Contributions}
%Please provide an author contributions statement using the CRediT taxonomy roles as a guide {\verb+\url{https://www.casrai.org/credit.html}+}.
Conceptualization: R.P; C.R.;T.F. Methodology: R.P; C.R.;T.F. 
%Data curation: A.C. Data visualisation: A.C. 
Writing original draft: R.P. All authors approved the final submitted draft.

%\renewcommand\bibpreamble{By default, this template uses \texttt{bibtex} and adopts the AMS referencing style. However, the journal you're submitting to may require a different reference style; specify the journal you're using with the class' \texttt{journal} option --- see lines 1--19 of \emph{sample.tex} for a list of options and instructions for selecting the journal}

% If using any of the following journal options:
%   wet, dap, dce, eds, prm, flw, jdm, psy, rsm
% then use the \printbibliography line instead of:
%\bibliography{literature}

\printbibliography

\end{Backmatter}
\end{document}